\documentclass{aa}
\usepackage{courier}
\usepackage{graphicx}
\usepackage{color}
\usepackage{txfonts}
\usepackage{amsmath}
\usepackage{enumitem}
\usepackage[colorlinks=true, citecolor=blue, urlcolor=magenta, linkcolor=darkgreen]{hyperref}
\usepackage[export]{adjustbox}
\usepackage{float}
\usepackage{subcaption}
\usepackage{natbib}
\definecolor{darkgreen}{rgb}{0, 0.42, 0.24}

\begin{document}

      \title{"Observations" of simulated dwarf galaxies}
   \subtitle{Star-formation histories from color-magnitude diagrams}
   \author{
                Shivangee Rathi \inst{1}
          \and
                  Michele Mastropietro \inst{1}
                  \and
                  Sven De Rijcke \inst{1}
                  \and
                  Carme Gallart \inst{4}
                  \and
                  Edouard Bernard \inst{2}        
                  \and
                  Robbert Verbeke\inst{3}
         }

   \institute{Department of Physics and Astronomy, Ghent University,
              Krijgslaan 281, 9000 Belgium\\
              \email{shivangee.rathi@ugent.be}
         \and
                Université Côte d'Azur, OCA, CNRS, Lagrange, F-06304 Nice, France
                \and
                 Institute for Computational Science, University of Zürich, Winterthurerstrasse 190, 8057, Zürich, Switzerland
                \and
         Instituto de Astrof\'{i}sica de Canarias, E-38200 La Laguna, Tenerife, Spain
             }

   \date{\today}

 \abstract
     {Apparent deviations between properties of dwarf galaxies from observations and simulations are known to exist, such as the "Missing Dwarfs" problem, the too-big-to-fail problem, and the cusp-core problem, to name a few. Recent studies have shown that these issues can at least be partially resolved by taking into account the systematic differences between simulations and observations.}
     {This work aims to investigate and address any systematic differences affecting the comparison of simulations with observations.}
     { To this aim, we analyzed a set of24 realistically simulated MoRIA ((Models of Realistic dwarfs In Action) dwarf galaxies in an observationally motivated way. We first constructed "observed" color-magnitude diagrams (CMDs) of the simulated dwarf galaxies in the typically used V- and I-bands. Then we used the CMD-fitting method to recover their star-formation histories(SFHs) from their observed CMDs. These solved SFHs were then directly compared to the true SFHs from the simulation star-particle data, mainly in terms of the star-formation rate(SFR) and the age-metallicity relation (AMR). We applied a dust extinction prescription to the simulation data to produce observed CMDs affected by dust in star-formation regions. Since future facilities, such as the James Webb Space Telescope and the European Extremely Large Telescope, will focus on the (near)-infrared rather than the optical, we also constructed and analyzed CMDs using the I- and H-bands.}
     {We find a very good agreement between the recovered and the true SFHs of all the simulated dwarf galaxies in our sample, from the synthetic CMD analysis of their V$-$I versus I as well as the I$-$H versus H CMDs. Dust leads to an underestimation of the SFR during the lastfew hundred million years, with the strength and duration of the effect dependent on the dust content. Overall, our analysis indicates that quantities like SFR and AMR derived from the photometric observations of galaxies are directly comparable to their simulated counterparts, thus eliminating any systematic bias in the comparison of simulations and observations.}
     {}

\maketitle


\section{Introduction}
Dwarf galaxies are the smallest and the most abundant type of galaxies in the Universe \citep{1994A&ARv...6...67F}. Being low-mass systems,they are more susceptible to the astrophysical processes that drive galaxy evolution than the massive galaxies and therefore serve as ideal systems to study the effect of these processes on their evolution. Numerical simulations based on the currently accepted $\Lambda$-Cold Dark Matter ($\Lambda$CDM) 
cosmological model \citep{2016A&A...594A..13P}, equipped with sub-grid models that  heuristically describe the gas-star interactions, are now able to provide predictions for the structural, dynamical, and stellar populations properties of dwarf galaxies across different galactic environments (\cite{2011MNRAS.416..601S, 2012A&A...538A..82R,2014ApJ...792...99S, 2014MNRAS.442.2909C,2015MNRAS.451.1247S, 2015MNRAS.454.1886S,2015MNRAS.454...83W, V15, V17} (hereafter referred to as V15 and V17, respectively), and \cite{2018MNRAS.476.3816F}).

A number of apparent mismatches between these predictions and the observations have challenged our understanding of (dwarf) galaxy formation and evolution. Two of the most notable issues are the intimately linked "Missing Dwarfs" problem and the so-called too-big-to-fail (TBTF) problem. The former is the mismatch between the predicted distribution of dwarf-sized dark-matter halos over circular velocity and the observed distribution of Local Group dwarf galaxies over rotation velocity \citep{1993MNRAS.264..201K,1999ApJ...522...82K,1999ApJ...524L..19M}. The TBTF problem
is the inability of simulations to reproduce the observed number densities of dwarf galaxies and their internal kinematics, both in the Local Group \citep{2011MNRAS.415L..40B} and in the field \citep{2016A&A...591A..58P}, at the same time. Moreover, dark-matter-only simulations predict that dark-matter halos should have a centrally cusped density distribution, whereas observations of dwarf galaxy rotation curves seem to prefer cored central densities (see e.g., \citet{2001MNRAS.325.1017V} and references therein). This is the cusp-core problem. Finally, we also wish to highlight the discrepancy between the observed and the predicted slope of the faint end of the(baryonic) Tully-Fisher relation \citep{2017MNRAS.464.2419S}.

Since the identification of these apparent shortcomings, it has become
clear that some of them can be at least partly solved by a proper
inclusion of baryonic physics in the simulations without abandoning
the $\Lambda$CDM model of cosmic evolution. For instance, dark-matter
cusps can be converted into cores by supernova induced gas motions
\citep{2012MNRAS.423..735C,2016MNRAS.459.2573R} if enough stars are
formed. Moreover, such outflows, and the dynamical response they
elicit from the dark matter, also lower the maximum circular
velocities of simulated dwarf galaxies, and this, in turn, alleviates the
Missing Dwarfs and TBTF problems \citep{2016MNRAS.457.1931S} without
fully solving them \citep{2017ARA&A..55..343B}.

It has been noted that systematic differences may exist between the
quantities being compared between simulations and observations,
leading to spurious mismatches between theory and observations. For
instance, \citet{2016A&A...591A..58P} point out that the TBTF
problem of field dwarfs can be explained if the circular velocities
derived from observed H{\sc i} kinematics are systematically smaller
than the actual circular velocities. Based on an analysis of the H{\sc
        i} kinematics of simulated dwarfs from the MoRIA (Models of Realistic dwarfs In Action) suite using the
same analysis codes also used by observers, V17 show that
these simulated dwarfs display 
exactly the dependence between observed
and actual circular velocities required to solve the TBTF of field
dwarfs. Likewise, \citet{2017MNRAS.466...63P} show that the gas
kinematics of simulated galaxies with centrally cusped dark-matter
distribution, when analyzed in the same way as observed H{\sc i}
kinematics, would preferentially lead to the retrieval of a cored
density distribution. In both cases, there is a marked difference
between the circular velocity as derived from the gas kinematics (a
quantity accessible by observations) and as derived from the mass
distribution (a quantity only accessible by simulators).

With the available wealth of resolved photometric data of nearby Local
Group dwarf galaxies, it has become possible to exploit their stellar
CMDs to study their individual stellar
populations and, in turn, to infer their star-formation histories (SFHs)
\citep{lcid3, 2014ApJ...789..148W, 2015ApJ...812..158M, 2016ApJ...823....9A,2017ApJ...837..102S}. This
is obviously crucial to our understanding of their formation and
evolution. As with the stellar and gaseous kinematics, it needs to be
checked whether systematic differences exist between stellar
population properties derived from simulations and observations.

In this paper, we continue our efforts in the direction of analyzing
dwarf galaxy simulations in an observationally motivated way (V15,
V17).  Assuming simulations as the ground truth, we re-construct the
SFHs of simulated dwarf galaxies from their CMDs using the synthetic
CMD method (e.g., \cite{Tosi1991}; \cite{Tolstoy&Saha1996};
\cite{Gallart1996a}; \cite{Dolphin1997}) used by observers, and see
how these results compare to simulations.

This paper is organized as follows. In Sect.
\ref{sec:simulations}, we list the main features of the MoRIA
simulations, which form the basis dataset of our study. Section
\ref{sec:CMD} describes in detail the construction of realistic mock
CMDs of simulated MoRIA dwarf galaxies from their simulation
star-particle data. In Sect. \ref{sec:SFH}, we give an outline of
the implementation of the synthetic CMD method, starting from creating
a synthetic CMD to the parameters crucial for the synthetic CMD method. In
Sect. \ref{sec:results}, we present and discuss the results based on
the I versus V$-$I CMDs. In the subsequent section, we do the same for
the I versus I$-$H CMDs. Lastly, we summarize and try to interpret our
results in Sect. \ref{sec:conclusions}.


\section{Simulations} \label{sec:simulations}
We use the MoRIA  suite of N-body/SPH simulations 
of late-type isolated dwarfs, presented in
\cite{V15}, as the primary dataset for this work. The MoRIA are high
resolution ($\mathrm{m_{bar} \sim 10^{3}\ M_{\odot}}$) dwarf galaxy
simulations performed with a modified version of the N-body/SPH-code
GADGET-2 \citep{gadget2}. The added astrophysical ingredients include:
radiative cooling, heating by the cosmic ultraviolet background 
radiation field, star formation, supernova and stellar feedback, 
and chemical enrichment including the Population-III stars \footnote{\tiny
        Simulations DG-18, DG-20, and DG-22 were simulated using a
        slightly different recipe, the major difference being the absence of
        Population-III feedback in these simulations, which causes them to have a
prominent early burst of star formation. }. More details on the
MoRIA suite of simulations are available in
\cite{DeRijcke2013}, \cite{Cloet-Osselaer2014}, \cite{V15},
\cite{Vandenbroucke2016}, and V17.

In this paper, we observationally analyze a set of 24 simulated MoRIA
dwarfs, with total stellar masses ranging between
$\mathrm{10^{6}\ M_{\odot}}$ - $\mathrm{10^{8}\ M_{\odot}}$. Table
\ref{table:1} presents some of the important characteristics of the
sample of dwarf galaxy simulations under study. For brevity, we
discuss the analysis of two representative
        simulations in the main body of the paper: DG-5, which has a star-formation rate (SFR) that rises with
        time, and DG-22, which formed most of its stars over ten billion
        years ago (see Table \ref{table:1}); we include the results of the other simulations in Appendix
        \ref{sec:AppA}. The following section details the construction of
realistic CMDs from the simulation star-particle data.

\begin{table}
\caption{Some of the basic properties of the MoRIA simulated dwarfs
  analyzed in this paper.} \label{table:1} \centering
\begin{tabular}{l l l l}     
\hline \hline
\vspace{0.05cm} \\
(1) & (2) & (3) & (4) \\
Simulation name & $\mathrm{\log_{10}(M_{*})}$ & $\mathrm{R_{eff}}$  & In V17  \\
{} & [$\mathrm{M_{\odot}}$] & [kpc] & \\
\vspace{0.05cm} \\
\hline
\vspace{0.05cm} \\
 DG-1 &           6.036 &     0.418 &    ... \\
 DG-2 &           6.326 &     0.364 &   ... \\
 DG-3 &           6.472 &     0.319 &    ... \\
 DG-4 &           6.488 &      0.246 &   ... \\
 DG-5 &           6.613 &     0.515 &   M-1 \\
 DG-6 &           6.718 &     0.263 &   ... \\
 DG-7 &           6.748 &     0.443 &   ... \\
 DG-8 &           6.852 &     0.420 &   M-2 \\
 DG-9 &           6.895 &     0.227 &   ... \\
DG-10 &           6.904 &     0.597 &   M-3 \\
DG-11 &           6.934 &     0.274 &   ... \\
DG-12 &           6.945 &     0.810 &   ... \\
DG-13 &            6.950 &     0.763 &   ... \\
DG-14 &           7.071 &     0.271 &    ... \\
DG-15 &           7.344 &     0.966 &   ... \\
DG-16 &           7.397 &     0.966 &    ... \\
DG-17 &           7.401 &      1.346 & M-4 \\ DG-18 & 7.496 & 1.000 &
... \\ DG-19 & 7.575 & 0.303 & M-5 \\ DG-20 & 7.680 & 1.150 &
... \\ DG-21 & 7.875 & 1.133 & M-6 \\ DG-22 & 7.912 & 0.964 &
... \\ DG-23 & 8.054 & 1.549 & M-7 \\ DG-24 & 8.556 & 1.686 & M-9 \\
\hline \hline
\end{tabular}
\tablefoot{Columns: (1) simulated galaxy; (2) the total stellar mass
  of the simulated galaxy; (3) the effective radius of the simulated
  galaxy based on the three-dimensional distribution of the star
  particles in it; (4) reference labels for MoRIA dwarfs previously
  discussed in V17.  }
\end{table}


\section{Constructing realistic color-magnitude diagrams} \label{sec:CMD}
Color-magnitude diagrams are a widely used tool to study the stellar fossil records of
nearby galaxies (\cite{lcid3}, \cite{Rubele2011}, \cite{Bernard2012},
etc.). This approach is currently limited to the local Universe, for
which resolved photometric data is accessible with the currently available
instruments. With the future space telescopes, like the James Webb
Space Telescope (JWST) \citep{2006SSRv..123..485G}, and upcoming ground-based
facilities, like the Thirty Meter Telescope
\citep{2015RAA....15.1945S}, the Giant Magellan Telescope
\citep{2014SPIE.9145E..1CB}, and the European Extremely Large Telescope (E-ELT)
\citep{EELT}, it will become possible to observe the resolved stellar
populations of galaxies outside the Local Group. This would open up
new environments to study galaxy CMDs.

First, we present a method to make a realistic CMD for a simulated
galaxy based on its simulation star-particle data. Codes that do exactly
this have already been presented in the literature. For instance,
\cite{DaSilva2012} present a fully stochastic code for synthetic
photometry that incorporates effects due to clustering, cluster
disruption, etc. However, we do not need to use the full might of such
codes for our goals, nor do we want to because they introduce physical
features that were not included in the original simulation. For our
purposes, it is sufficient to sample individual stars from the stellar
particles and obtain their photometric magnitudes from the closest
matching isochrone in a stellar evolution library.

To obtain the CMD, we centered the snapshot on the center of
        mass of the star particles and rotated it in the face-on
        orientation. As a pre-selection step, we selected only the
star particles lying within 1 kpc of the galaxy center and excluded the
extremely metal-poor Population-III star particles (with [Fe/H] <
-5). In addition to the star-particle data (i.e., a star particle's
mass, age, metallicity, and alpha abundance), we used two other
ingredients to construct CMDs of simulated dwarf galaxies: (i) an
initial mass function (IMF) and (ii) a stellar evolutionary library,
or, more specifically, the stellar isochrones. We discuss these in more detail in the following sub-sections, followed by a detailed
description of the steps involved in constructing a realistic
"observed" CMD of a simulated galaxy. To avoid confusion, the terms
"star particle" and "star" are used for a simulation star particle and
an individual star, respectively.

\subsection{Initial mass function} \label{ssec:IMF}
The dwarf galaxy simulations have star particles with known ages,
masses, metallicities, and alpha-abundances. Each star particle is
treated as a simple stellar population (SSP). From each star particle,
we sampled several individual stars assuming that the distribution
of the sampled stars follows a Chabrier IMF \citep{chabrier2003}:

\begin{equation} \label{eqn:imf}
\begin{split}
\frac{d\phi(m)}{dm} & =
        \begin{cases}
                \displaystyle \frac{
                  \exp[{-\mathrm{A}^{2}(\log_{10}{m} +
                      \mathrm{B})^{2}}]}{m\ \mathrm{C}} & m \leq 1
                \mathrm{M_{\odot}} \\\\ \displaystyle m^{-2.3} & m > 1
                \mathrm{M_{\odot}} \\
        \end{cases}
\end{split}
, \end{equation}

\hspace{0pt} where $ A^{2} = 1.0502,\ B = 1.1024,\ \mathrm{and} \ C =
e^{-(AB)^{2}} $. We sampled stars with masses between $
\mathrm{0.1\ M_{\odot}} $ and $ \mathrm{70\ M_{\odot}} $. Integrating
the IMF in Eq. \eqref{eqn:imf} within the these stellar mass
limits, and normalizing, gives:

\begin{equation} \label{eqn:cum_imf}
\begin{split}
        \Phi( \mathrm{M}) & = \frac{1}{\mathrm{N}}\ \int
        \limits_{0.1\ \mathrm{M_{\odot}}}^{\mathrm{M}} { \ d \phi(m) }
        \\\\ & = \begin{cases} \displaystyle
          \frac{\Phi_{1}(\mathrm{M})}{\mathrm{N}} & \mathrm{M \leq 1
            M_{\odot}} \\\\ \displaystyle \frac{\Phi_{1}(1) +
            \Phi_{2}(\mathrm{M})}{\mathrm{N}} & \mathrm{M > 1
            M_{\odot}}
                \end{cases}
\end{split}
, \end{equation}
where

\begin{gather*}
\Phi_{1}(\mathrm{M}) = \mathrm{D\ [\ \mathrm{erf}(A (\log_{10}{M} +B))
    - \mathrm{erf}(A(B-1))\ ]} \quad \textrm{and}\\
\\
\Phi_{2} (\mathrm{M}) = \frac{1-\mathrm{M}^{-1.3}}{1.3}
\textrm{,} \quad \textrm{with}\\
\mathrm{D} = \frac{\sqrt{\pi}}{\mathrm{2AC} \log_{10}{e}} \textrm{,}
\quad \mathrm{erf(x)} = \frac{2}{\sqrt{\pi}} \int\limits_{0}^{x}
e^{-t^{2}}\ dt,\ \textrm{and}\ \\ \mathrm{N} = \Phi_{1}(1) +
\Phi_{2}(70).
\end{gather*}

Inverting Eq. \eqref{eqn:cum_imf} gives the stellar mass as a
function of normalized cumulative IMF, $\mathrm{\Phi}$:

\begin{equation} \label{eqn:inverse_imf}
\begin{split}
M(\Phi) & =
        \begin{cases}
                \displaystyle 10^{\big{ \{ } \mathrm{erfinv} \big[
                    \frac{\Phi\ N}{\mathrm{D}}
                    +\ \mathrm{erf(A\ (B-1))} \big] - \mathrm{B} \big{
                    \} } /\mathrm{A}} & \Phi \leq
                \Phi_{1}(1)\\\\ \displaystyle \big[ 1- 1.3\ ( N\ \Phi
                  - \Phi_{1}(1)) \big]^{-1/1.3} & \Phi > \Phi_{1}(1)
        \end{cases}
\end{split} \\
.\end{equation}

\begin{figure}
 \centering
 \includegraphics[width=\hsize]{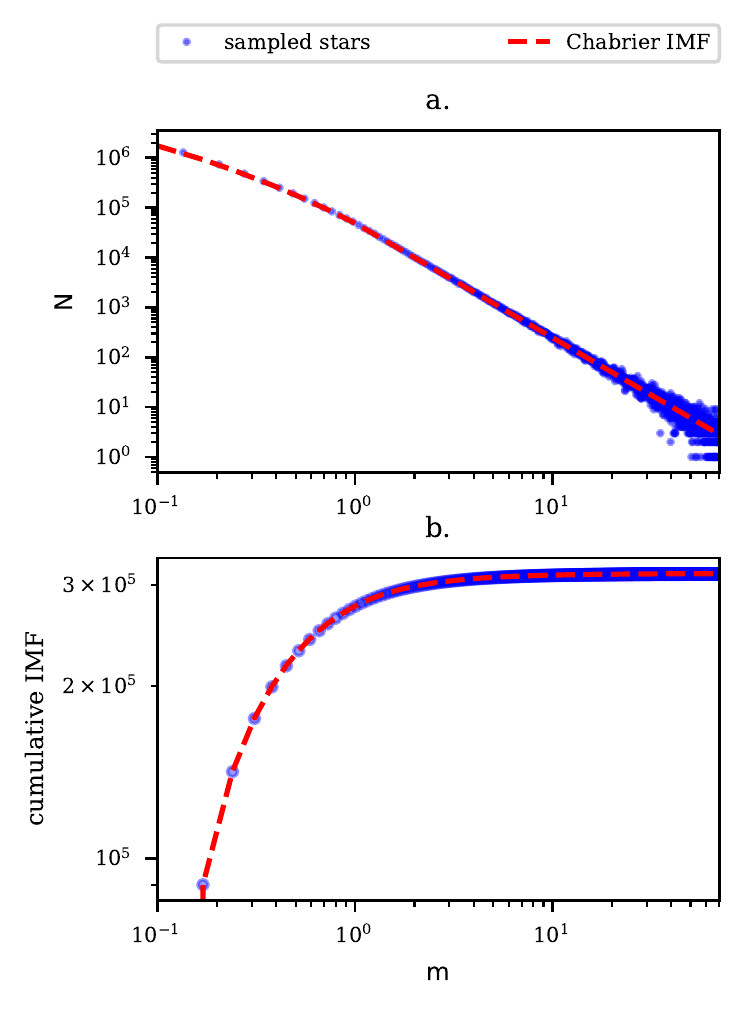}
    \caption{Mass distribution of the sampled stars of the simulated
        dwarf galaxy DG-5 based on the method described in Sect.        \ref{ssec:IMF}. Panel (a) shows the Chabrier IMF described by
        Eq. \ref{eqn:imf} and panel (b) shows the cumulative IMF
        given by Eq. \ref{eqn:cum_imf}. The sampled distributions are represented by blue circles and the analytical functions are shown by the dashed red lines.}
    \label{fig:IMF}
\end{figure}

Therefore, if we randomly draw numbers from [0,1) for the normalized
quantity $\mathrm{\Phi (M)}$, then, with the help of Eq.
\eqref{eqn:inverse_imf}, we can sample the masses (M) of the
individual stars from a star particle until the total mass of the
star particle is reached. However, the mass of the star particle is
never sampled exactly, and, in the present work, we allowed for
over-sampling by one star. This is similar to the "stop after"
sampling discussed in \cite{Haas2010}, where it is discussed in the
context of sampling stars from a star cluster. This over-sampling by
one star results in an error, which is the difference between the actual mass of the star particle and the total mass of the stars
sampled from it. We find a maximum over-sampling error of $
\mathrm{0.12\ \%}$ of the total mass of all star particles. Figure
\ref{fig:IMF} shows the distribution of all the sampled stars for
the simulated galaxy DG-5 in panel (a) and its
cumulative in panel (b), compared to their analytical forms given by
Eqs. \ref{eqn:imf} and \ref{eqn:cum_imf}, respectively.

\subsection{Stellar evolution library}\label{ssec:isochrones}
Once all the star particles are sampled into individual stars,
one can, with the knowledge of a star's mass, age, metallicity, and
alpha-abundance, derive its photometric magnitudes with
the help of the stellar evolution libraries. In particular, we used the
BaSTI (Bag of Stellar Tracks and Isochrones) \footnote{\tiny
        \url{http://basti.oa-teramo.inaf.it/}} stellar evolution library to
obtain the V- and I-band magnitudes of the sampled stars to construct
their typically used V$-$I versus I CMDs. The sampled stars cover a broad
range of masses ($\mathrm{0.1\ M_{\odot}}$-$\mathrm{70\ M_{\odot}}$);
therefore, to cover the maximum range of masses of the
sampled stars, we used a combination of two complementary stellar
evolution models (canonical) within BaSTI: (i) the "standard model"
\citep{Pietrinferni2004, basti2006, basti2013} and (ii) the model
extending to the very-low-mass (VLM) stars (based on \cite{bastiVLM}).

The standard model comprises both scaled-solar and
alpha-enhanced asymptotic giant branch (AGB) extended isochrones with
an AGB mass loss efficiency parameter, $ \mathrm{\eta = 0.4} $. The
isochrones in the standard model follow the stellar evolution from
the pre-main sequence to the early-AGB phase. They cover a range in
mass from $ \mathrm{0.5\ M_{\odot}}$ to $\mathrm{10\ M_{\odot}} $,
while covering a broad range in metallicity $ \mathrm{10^{-5} \leq Z
        \leq 0.05 }$.

On the other hand, the VLM model, as the name suggests, covers the
low-mass end of the sampled stars. The VLM model has the scaled-solar
isochrones with an AGB mass loss efficiency parameter, $ \mathrm{ \eta
        = 0.4} $. The isochrones in the VLM model cover the hydrogen-burning
stars going from the faint end of the main sequence up to the main
sequence turnoff. They extend to VLM stars with masses down
to $ \mathrm{0.1\ M_{\odot}}$ (and up to $ \mathrm{2\ M_{\odot}} $) and
cover metallicities in the range $ \mathrm{2 \times 10^{-4} \leq Z
        \leq 2 \times 10^{-3} }$. Both models cover a range in age from
0.03 Gyrs to 14 Gyrs.

\begin{figure}
 \centering
 \includegraphics[width=\hsize]{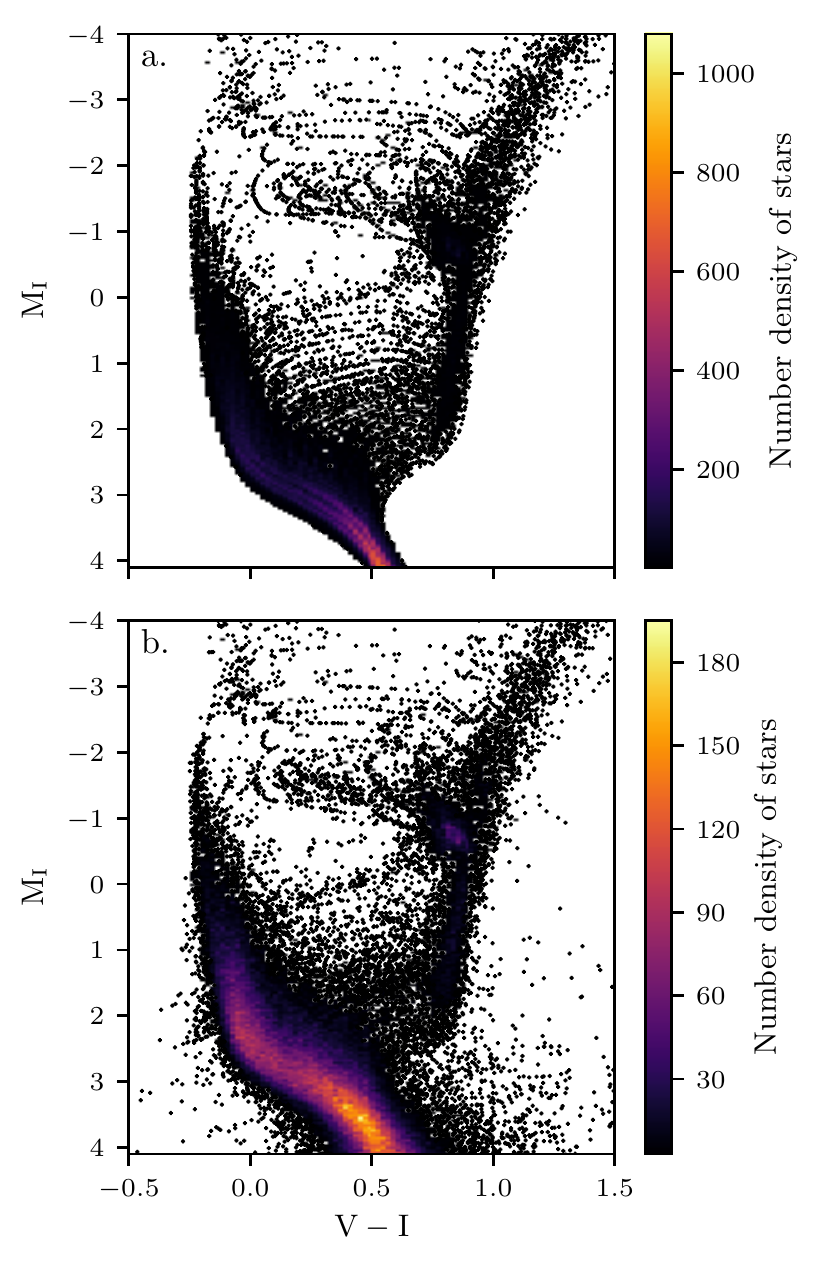}
    \caption{Hess representation of un-convolved and convolved
      versions of CMD of the simulated dwarf galaxy DG-5. The
      magnitudes in panel (a) are a direct result from interpolating
      on the best matching isochrones (based on the age, metallicity,
      and mass of the sampled stars). The magnitudes in panel (b) are
      obtained by simulating observational errors using the crowding
      tables used in \cite{Meschin2014} to mimic the quality of real
      observational data. The color bars indicate the number density
      of stars in the corresponding plots, and the sparse black points
      represent individual stars in the CMD.}
    \label{fig:CMD}
\end{figure}

\bigskip
\begin{table*}[hbt!]
\caption{\label{table:mCMD} Parameters used for creating a
  synthetic CMD.}  \centering
\begin{tabular}{lc}
\hline\hline
\vspace{1pt} \\ Parameter & Value\\
\vspace{0.000cm} \\ \hline
\vspace{0.005cm} \\ Stellar evolution library & BaSTI stellar
evolution library \text{ \citep{Pietrinferni2004}} \\ Bolometric
correction library & \text{\cite{Castelli&Kurucz2003}}\\ Number of
synthetic stars & 1 million \tablefootmark{a}\\ RGB and AGB mass loss
parameters, respectively & 0.35, 0.4 \\ Star-formation rate (SFR(t)) &
Constant star formation between 0 and 14 Gyrs.\\ Metallicity law Z(t)
& Upper and lower metallicity laws \tablefootmark{b}\\ Initial mass
function & Chabrier IMF \tablefootmark{c}\\ Binary fraction & 0
\\ \hline \hline
\end{tabular}
\tablefoot{ \\ \tablefoottext{a}{Due to long run-times, IAC-star
    allows for the calculation of a maximum of 1 million synthetic stars at a
    time. So we generated 50 different 1-million synthetic CMDs and
    combined them to obtain a synthetic CMD with 50 million stars.}
  \\ \tablefoottext{b}{This gives a uniform dispersion in metallicity
    between 0.0001 and 0.03.} \\ \tablefoottext{c}{We computed stars
    with masses between 0.1 $\mathrm{M_{\odot}}$ and 70
    $\mathrm{M_{\odot}}$. IAC-star only allows for a power-law form of
    the IMF, so we approximated the low-mass end of the Chabrier IMF
    ($\mathrm{ 0.1\ M_{\odot}} \leq m \leq \mathrm{1\ M_{\odot}}$)
    with a power law of the form $ \phi (m)\ dm = m^{-1.6}$. } }
\end{table*}

For each of the star particles of a simulated galaxy, based on the
age, metallicity, and alpha-abundance of a star particle, a best
matching isochrone was selected from the broad range of isochrones
described above. In cases where the star particle lies beyond the
age-metallicity bounds of the set of available isochrones, such as the few extremely metal-rich (Z>0.03) and/or 
very young star particles, it is characterized by the closest limiting
isochrone. For example, a young metal-rich star with $\mathrm{Z \geq
        0.05 }$ and $ \mathrm{age \leq 0.03\ Gyrs}$ will be approximated by
the closest available isochrone with Z = 0.03 and age = 0.03
Gyrs. Now, based on the mass of a sampled star, its photometric
magnitude was derived by linearly interpolating in the
magnitude-$\mathrm{\log(mass)}$ plane \footnote{ \tiny Due to the near
        power-law form of the luminosity-mass relation, the
        magnitude-$\mathrm{\log(mass)}$ interpolation yields a good
        approximation.} between the magnitudes of its closest mass
neighbors. This was done on a star-by-star basis for all the stars
sampled from all the star particles of a simulated galaxy. The
resulting CMD of a simulated galaxy needs to be convolved with
observational errors to make the analysis observationally compliant.

We simulated observational errors on the resulting CMD using a
  procedure similar to that discussed in \cite{Gallart1996a} and
  \cite{Hidalgo2011}, where it is used to introduce observational
  errors in the synthetic CMD. This method is based on the artificial
  star tests, wherein artificial stars with known magnitudes are
  injected into the observed frames and the comparison of the injected
  and recovered stars provides information on crowding and
  incompleteness. For our purposes, we are only concerned with the
  incompleteness. Any star that was not recovered was discarded from the
  CMD. 
  Furthermore, for the stars that were recovered, the
  observational errors were calculated as the difference between the
  recovered and injected magnitudes of the artificial star. The
  information about incompleteness and observational errors obtained
  from the artificial star tests are given in crowding tables.  In
  particular, we used the crowding table used in \cite{Meschin2014} to
  simultaneously simulate the incompleteness as well as the
  observational errors in the mock CMDs. In this way, observational
  errors were introduced in both the mock CMDs of the simulated
  galaxies and the model CMD.

The mock CMDs of the simulated galaxy DG-5 are shown in Fig.
\ref{fig:CMD}. The top panel shows the CMD that results directly from
the interpolated magnitudes; the CMD in the bottom panel has
been, in addition, convolved with the magnitude dependent photometric
errors and is, therefore, a more realistic representation of an
observed CMD. From panel (b) of Fig. \ref{fig:CMD}, it can be noted
that the isochrone tracks are still visible for the brightest stars,
where the photometric uncertainties are minimal. This is inherent to
the process of obtaining photometric magnitudes of individual stars
sampled from the same stellar particle. Each stellar particle
constitutes a single stellar population with a given age and
metallicity, and the photometric magnitudes of all the stars sampled
from it are, therefore, obtained from a single isochrone. Since the
synthetic CMD method relies on counting the number of stars in
color-magnitude bins, which in practice smoothes out the distribution
of stars in the CMD (see \cite{iac-pop}), this feature does not
influence our results. Having discussed the construction of realistic CMDs from simulation
star-particle data, in the next section we discuss the method to
retrieve the SFHs from the mock CMDs.


\section{Star-formation history from the mock CMDs} \label{sec:SFH}
The star-formation history (SFH) is crucial to understanding the evolution of a galaxy, as it
gives important clues about the internal and external drivers of its
evolution. The SFH of a complex stellar population, such as a galaxy,
can be inferred from its CMD by comparing the observed CMD with a
theoretical or model CMD, encompassing various possible scenarios of
its evolution. This is known as the synthetic CMD method. To put
it simply, the synthetic CMD method is based on fitting the number of
stars in various regions of the observed CMD with SSPs underlying the respective regions in a model
CMD. We used the CMD-fitting technique, in particular the method
described in \cite{iac-pop} as implemented in \cite{Bernard2015a}, to
retrieve the SFH of the simulated dwarf galaxies from their mock
CMDs. In the following section, we briefly review the parameters involved in
the making of a model CMD, highlight the key features of the
synthetic CMD method, and finally discuss the implementation of the
synthetic CMD method used in this work.

\subsection{The model CMD} \label{ssec:mCMD}

To create the synthetic CMD, we used the publicly available
IAC-star \footnote{ \tiny \url{http://iac-star.iac.es/cmd/index.htm}}
code \citep{iac-star}. This code relies on a number of input parameters,
which, along with their values used in this work, are summarized in
Table \ref{table:mCMD}. It must be noted that all the parameters
described in Table \ref{table:mCMD} are kept as close as possible to
those used in the construction of mock CMDs from the simulation
star-particle data to avoid any systematic differences in the
subsequent results. We constructed a single synthetic CMD with 50
million stars, which is used for analyzing all the simulations (the
reasons behind this choice are discussed in Sect.
\ref{ssec:solveSFH}). Finally, incompleteness and other observational
errors were simulated in this synthetic CMD in the same way as was
described in Sect. \ref{sec:CMD} for the case of mock CMDs. The
resultant CMD was then directly comparable to the mock CMDs and is
referred to as the model CMD.

\subsection{The synthetic CMD method}\label{ssec:solveSFH}
In the mock CMDs, the total number of sampled stars varies from
simulation to simulation and ranges from a few hundred thousand stars to a few
million stars. Ideally, depending on the number of stars in the
mock CMDs, each would require a different model CMD with
a sufficient number of stars. For example, a mock CMD with a million
stars would ideally require a model CMD with $\mathrm{\sim}$0.2
billion stars for a reasonable comparison. \linebreak

However, making, as well as handling, such model CMDs with millions
of stars is computationally very expensive. Therefore, we instead
limited the number of stars in the densely populated mock CMDs to
the population of 200,000 randomly selected stars and used a single model
CMD for analyzing all our simulations. This 200,000 star threshold is
motivated by observational studies such as those in the LCID
project\footnote{\tiny
        \url{http://www.iac.es/proyecto/LCID/?p=home}}, \cite{bernard2018},
etc., where the authors report similar numbers of observationally
resolved stars for nearby systems. Furthermore, this not only saves the
computational expense, but also ensures that the model CMD always has
a lower Poisson noise than the mock CMD.

With the mock CMDs of simulated dwarf galaxies and a comparable model
CMD in hand, we used the aforementioned implementation of the
synthetic CMD method to solve for the SFHs of the simulated dwarf
galaxies from their mock CMDs. This implementation of the CMD-fitting
method has the provision to find 
the color-magnitude offset between
the observed and the model CMDs, among other tunable input
parameters. Since both the mock and the model CMDs are based on the
same stellar evolutionary library, we set this offset to zero. Another
interesting parameter is the selection of color-magnitude regions
(referred to as "bundles" in \cite{iac-pop}) based on which
solution SFH is computed. \cite{Ruiz-Lara2018} show that
including as many evolutionary phases as possible in the bundles
leads to a more reliable recovered SFH, even though some of the
evolutionary phases (such as the red-giant branch phase) 
might be affected by larger uncertainties. Consequently, in our analysis, we used a single
bundle covering the entire model CMD. The bundle was further
sub-divided into "bins," and the solution SFH was computed based on the
comparison between the mock and the model
CMDs of the stars in these bins. We set the bin size in the defined bundle to 0.025
$\mathrm{\times}$ 0.2 col$-$mag.
 
With the input parameters set to suitable values, a minimization
algorithm tries to fit the number of stars in each of the bins in
the mock CMD with the SSPs underlying the respective bins on the
model CMD. We used a set of $\sim$350 SSPs \footnote{\tiny{Grid formed
    by the following age and metallicity bins. Age (Gyrs): [0., 0.5,
      1., 1.5, 2., 2.5, 3., 4., 5., 6., 7., 8., 9., 10., 11., 12.,
      13., 14.]. \linebreak Metallicity (Z): [0.0001, 0.00015, 0.0002,
      0.0003, 0.0004, 0.0007, 0.0011, 0.0016, 0.0022, 0.0029, 0.0037,
      0.0046, 0.0056, 0.0067, 0.0079, 0.0092, 0.0106, 0.0122, 0.0140,
      0.0160, 0.0182, 0.03].}} for our analysis. The goodness of the
fit was measured by the Poisson equivalent of $\mathrm{\chi^{2}}$:
$\mathrm{\chi^{2}_{P}}$, adopted from \cite{Dolphin2002}. The
coefficients of the best-fitting solution CMD are directly
proportional to the birth-mass of the corresponding SSPs.


\section{Results and discussion} \label{sec:results}

\subsection{V$-$I versus I CMDs}

\begin{figure*}
        \centering
        \begin{subfigure}[t]{0.5\textwidth}
                \centering
                \caption*{(i) SFH of DG-5}
                \includegraphics{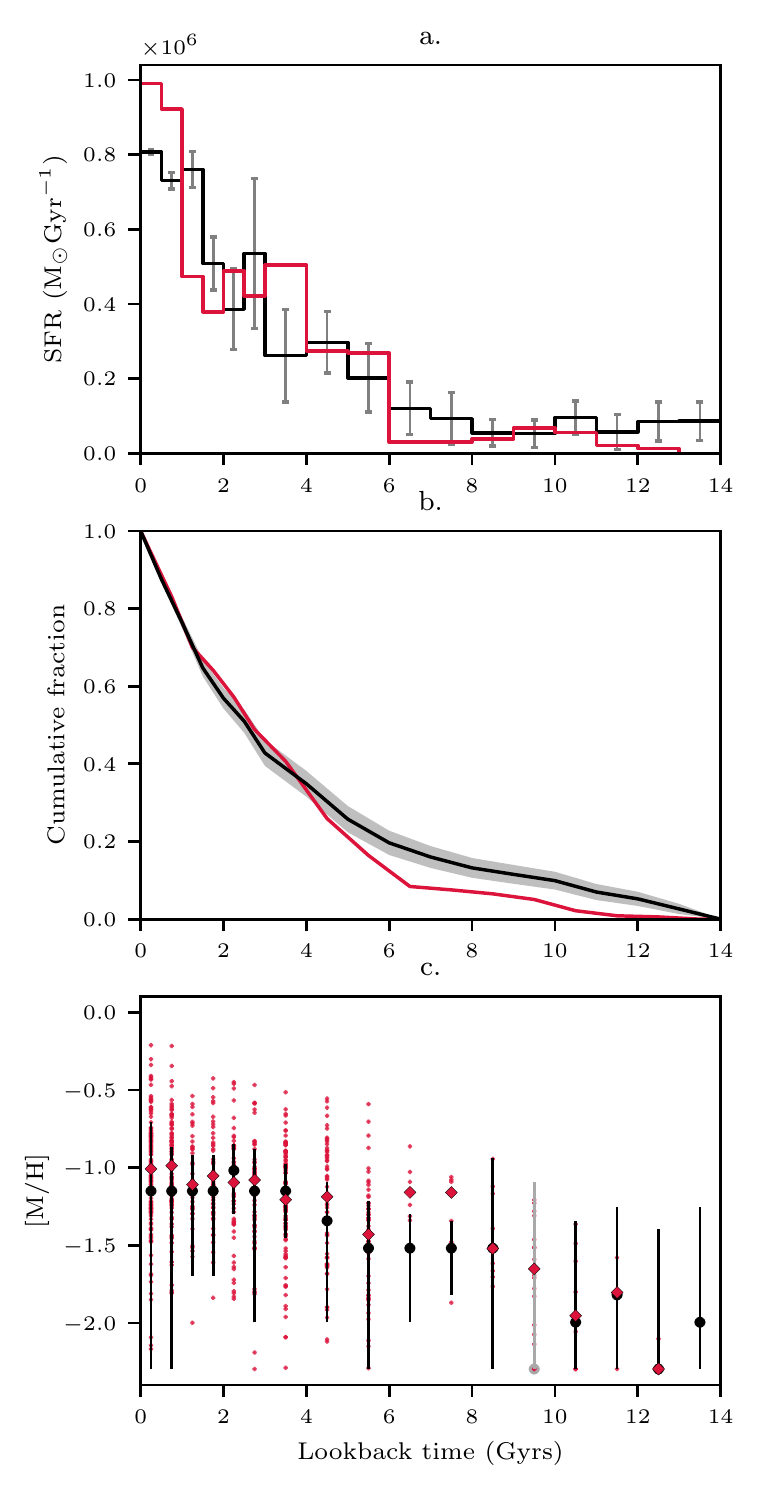}
        \end{subfigure}%
        ~
        \begin{subfigure}[t]{0.5\textwidth}
                \centering
                \caption*{(ii) SFH of DG-22}
                \includegraphics{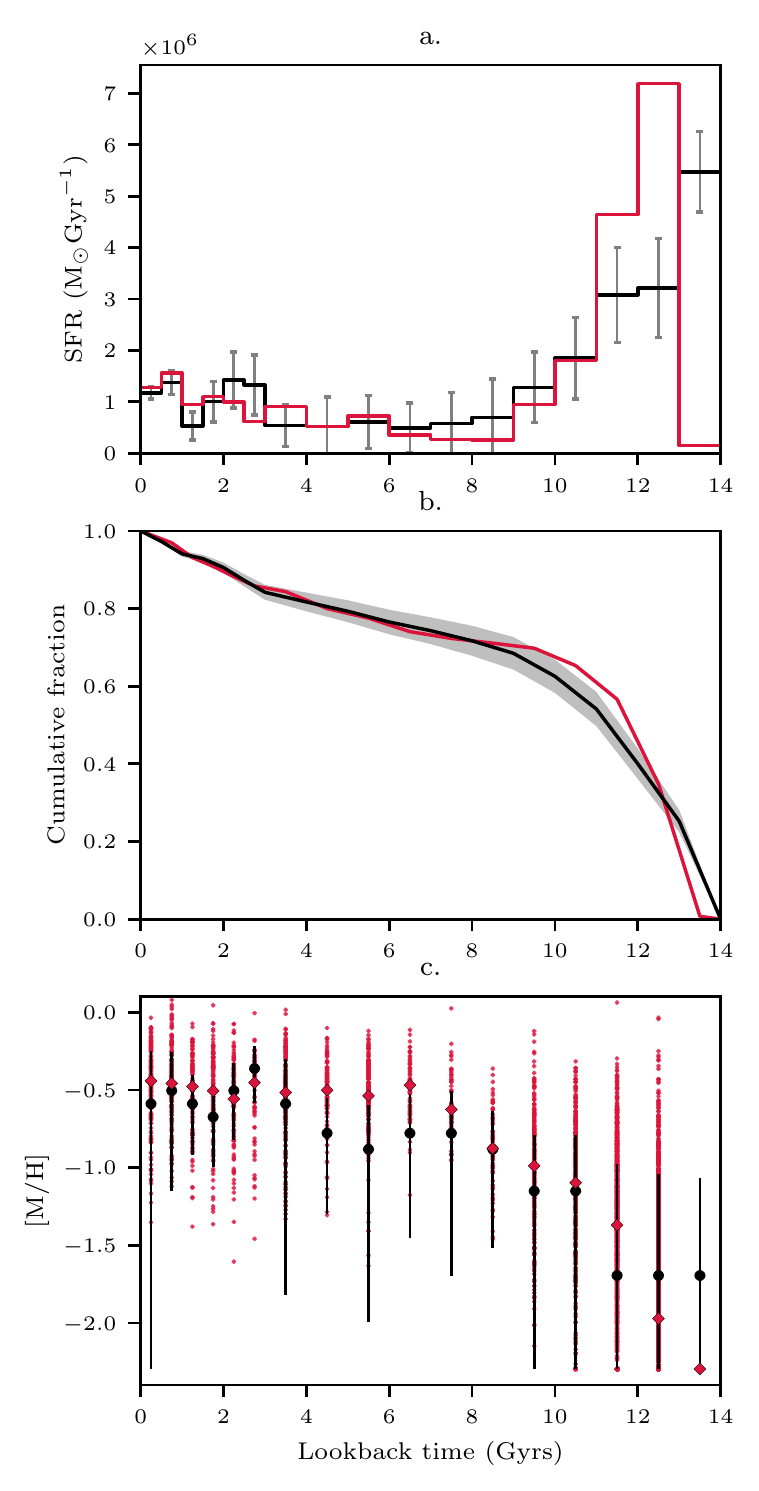}
        \end{subfigure}
        \caption{Results from the V$-$I vs. I CMD analyses of
            DG-5 (left) and DG-22 (right): solved SFH from the
          synthetic CMD method using 357 SSPs (in black) compared with
          the simulation star-particle data (in red) for the simulated
          dwarf galaxies DG-5 and DG-22 (see Table \ref{table:1} for
          more details). Panels (a) show the SFR as a function of the
          lookback time; panels (b) show the cumulative mass fraction; and
          panels (c) show the AMR. The gray points in the AMR depict the
          age bins where less than 1$\mathrm{\%}$ of the total
          star formation took place, the black and gray lines show the
          error on the solved metallicity, and the scattered red points
          in the AMR show the true metallicities of each of the
          star particles in various age bins.}
        \label{fig:sfh_DG5_DG22_VI}
\end{figure*}

The SFR and the age-metallicity relation (AMR)
are the two main results from the synthetic CMD method. Quantities
from the solved SFHs 
are compared directly to their true values from
the simulation star-particle data. Such a comparison for
the selected representative cases, DG-5 and DG-22, is
        shown in Fig. \ref{fig:sfh_DG5_DG22_VI}, where the top panel shows
        the comparison of the SFR, the middle panel
        shows the comparison of the cumulative SFR (or
        cumulative fraction), and the bottom panel shows the comparison of
        the AMR. Red and black colors correspond
to the true and the solved quantities, respectively. These are
discussed in more detail in the following section. In addition, Fig.
\ref{fig:hessComp_DG5_DG22_VI} shows the mock CMD (panel a), the
solution CMD (panel b), and their likeness (panels c and d) for
DG-5 and DG-22. The results of the other simulated galaxies are presented in
Appendix \ref{sec:AppA}. 

\begin{figure*}
        \centering
        \begin{subfigure}[t]{1\textwidth}
                \centering
                \caption*{(i) The observed/mock and solution CMDs of DG-5}
                \includegraphics[width=\hsize]{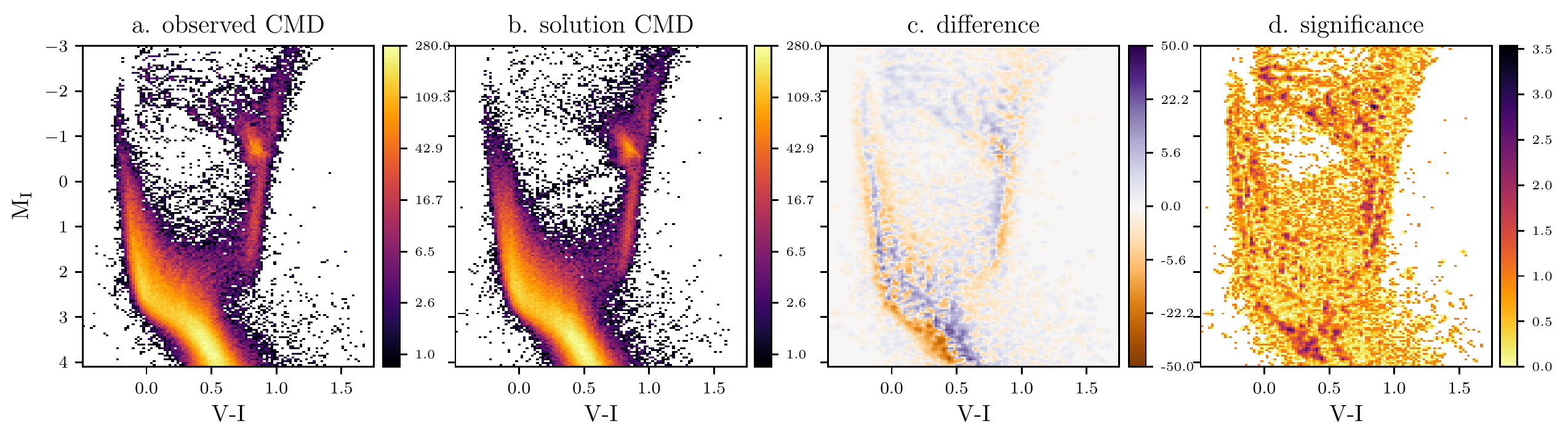}
        \end{subfigure}%
        
        \begin{subfigure}{1\textwidth}
                \centering
                \caption*{(ii) The observed/mock and solution CMDs of DG-22}
                \includegraphics[width=\hsize]{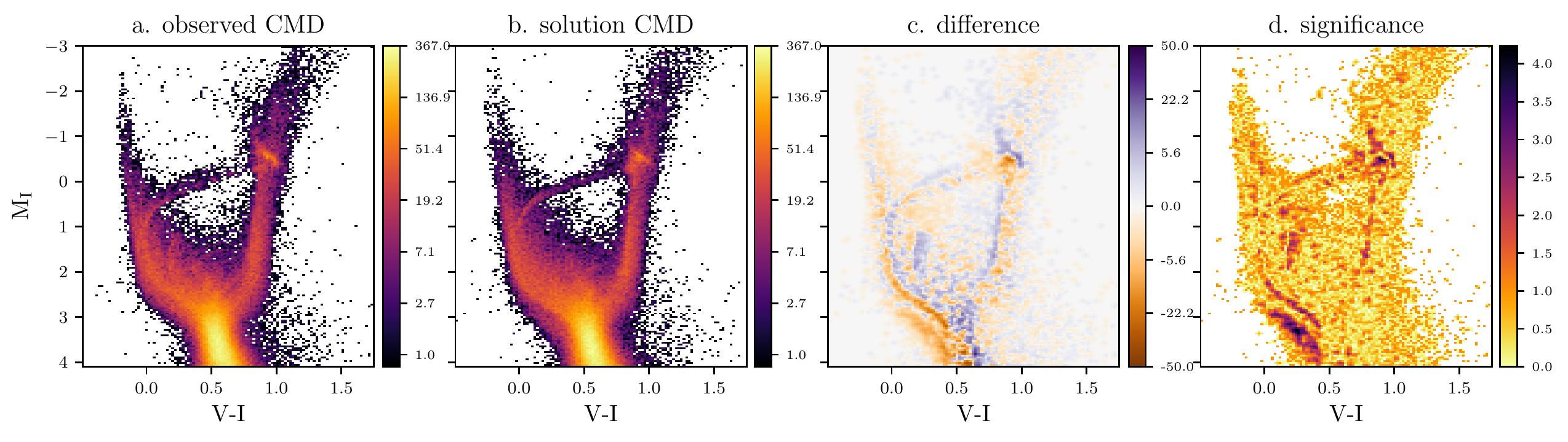}
        \end{subfigure}
        \caption{The comparison of mock CMDs with solution
            CMDs of DG-5 (top) and DG-22 (bottom): As indicated, the
          first column represents the mock or observed CMDs, the
          second column represents the solution CMDs, the third column represents the
          residuals, and the fourth column represents the significance for the
          simulated dwarf galaxies DG-5 (upper panel) and DG-22 (lower
          panel). }
\end{figure*}
 
The solved SFR is
directly compared to the true SFR from the simulation star-particle
data.  Panel (a) in Fig. \ref{fig:sfh_DG5_DG22_VI} shows the comparison of the true and the recovered SFR for the simulated dwarf
galaxy DG-5. We find a good agreement between the true and the
recovered SFRs in all of the simulated dwarf galaxies in our
sample. Notably, the star-formation peaks are well-recovered with
a fair constraint on the time and duration of the star-formation
phases. 

{}
 The cumulative SFH, or cumulative
  mass fraction, is defined as the fraction of a galaxy's total stellar
  mass that has been formed up to a certain time in its history since
  the birth of its first star. By definition, the cumulative fraction
  is zero before the first star is born and should be one at the
  current time. Due to the accumulative nature of the cumulative
  fraction, the errors on the solved SFR cannot be directly translated
  into the errors on the cumulative fraction. We therefore used
  bootstrapping to assign reasonable errors to the recovered
  cumulative fraction: From the solved SFR and its associated error in
  each age bin, we sampled 10,000 normally distributed alternative
  SFRs. Then we took the mean of the 10,000 different cumulative
  fractions resulting from the 10,000 different SFRs.  Finally, we
  calculated the error on the cumulative fraction as the root-mean-square deviation
  on the mean cumulative SFH. Panel (b) in Fig.
  \ref{fig:sfh_DG5_DG22_VI} shows the comparison of the true and the
  solved cumulative fraction of the simulated dwarf galaxy DG-5 as a
  function of the lookback time.
   
  Due to the prevalence
of various definitions of metallicity in the literature, it is
important that we specify the definition used in this work. The
following definition of metallicity was used in all our calculations:

\begin{gather} \label{eqn:metallicity}
\mathrm{ \Bigg[ \frac{M}{H}\Bigg] = log_{10}\Bigg(
  \frac{Z}{Z_{\odot}}\Bigg) },
\end{gather}
where Z is the mass fraction comprised of the total metal content of a
star, and $\mathrm{Z_{\odot}\ =\ 0.0198}$ \citep{Grevesse&Noels1993},
is its solar value.

In each age bin, the true metallicity from the star-particle data was
calculated as the median of the metallicities of the star particles
born in that age bin. Panel (c) in Fig. \ref{fig:sfh_DG5_DG22_VI}
shows the comparison of the true and the recovered AMR for the
simulated dwarf galaxy DG-5. The gray markers denote the age bins
where the solved SFH indicates that less than 1$\mathrm{\%}$ of the total
star formation took place and are thus deemed unfit for comparison; the
red dots show the scatter of metallicity of the individual
star particles in each of the age bins. We also find a good agreement
between the true and the recovered AMRs.


\subsubsection{Radial dependence of the retrieved SFH }

\begin{figure}[h]
 \centering \includegraphics[width=\hsize]{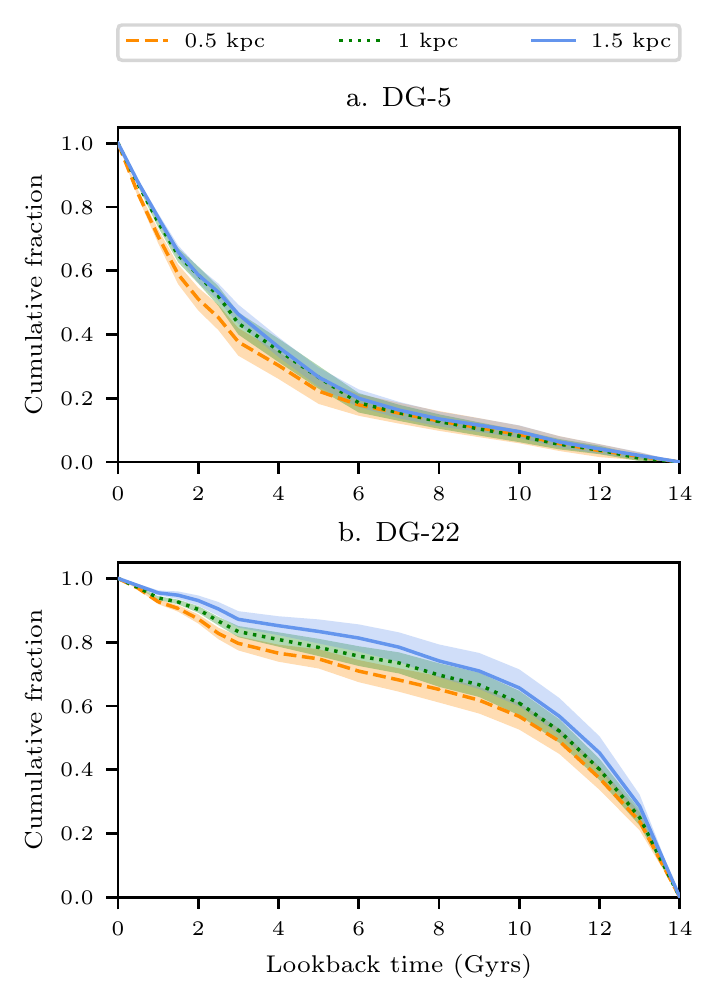}
    \caption{ Effect of using different apertures on the
                retrieved cumulative mass fraction of DG-5 (top) and DG-22
                (bottom): The above plot shows a comparison of the cumulative
                SFRs from considering three different apertures, where blue
                (solid), green (dotted), and orange (dashed) lines represent the
                solved SFR from considering 1.5 kpcs, 1.0 kpc, and 0.5 kpc radial
                apertures from the center of the galaxy, respectively.} 
    \label{fig:sfh_vs_r_DG5}
\end{figure}

In addition to the hitherto studied inner 1 kpc region of the
simulated galaxies, we also studied the effect of using a smaller (0.5
kpc) and a larger (1.5 kpc) aperture on the resulting SFH. Results from this analysis for DG-5 and DG-22
  are shown in Fig. \ref{fig:sfh_vs_r_DG5}. We observe that the
    cumulative mass fraction approaches its maximum value more
    steeply at small lookback times as we go to smaller apertures. In
    other words,~the fraction of young stars, and hence the rate of
    recent star formation, increases as one goes to smaller apertures.
   This age gradient is in line with stellar population studies of
  the Local Group dwarf galaxies (\cite{deBoer2012},
  \cite{Battaglia2012}), where a similar radial age gradient is
  observed.

\subsubsection{Effects of dust extinction}

\begin{figure*}
        \centering
        \begin{subfigure}[t]{1\textwidth}
                \centering
                \caption*{(i) Effect of dust extinction on the CMD of DG-5}
                \includegraphics[width=\hsize]{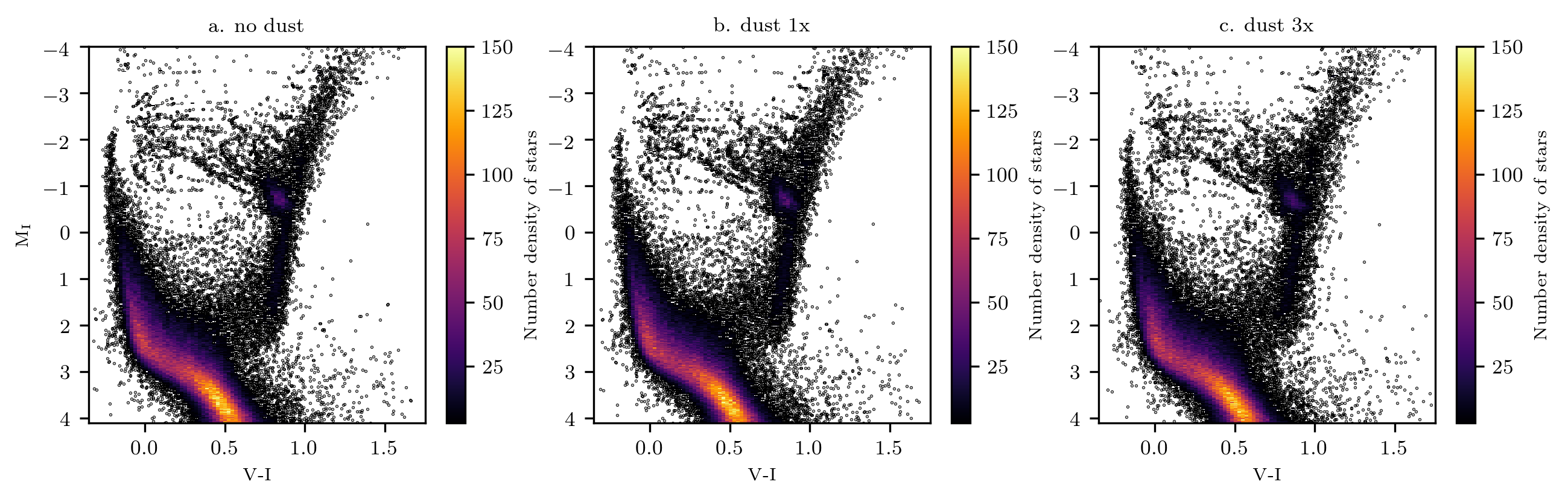}
        \end{subfigure}%
        
        \begin{subfigure}{1\textwidth}
                \centering
                \caption*{(ii) Effect of dust extinction on the CMD of DG-22}
                \includegraphics[width=\hsize]{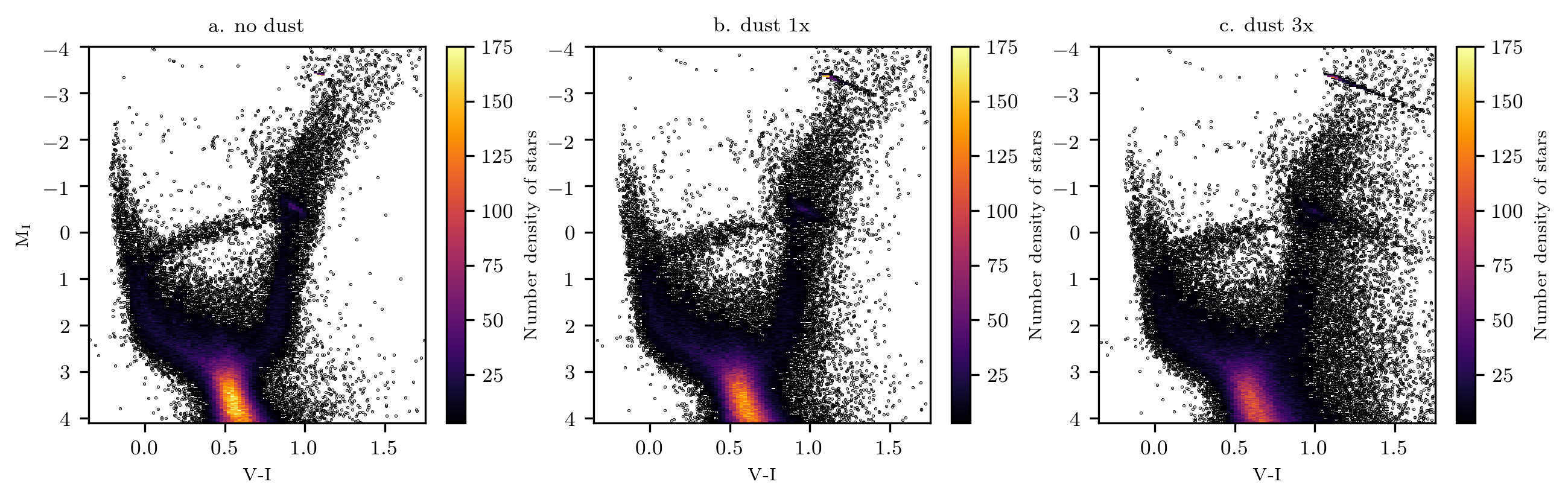}
        \end{subfigure}
        \caption{Effects of the adopted extinction prescription on
                        the observed CMDs of DG-5 (top) and DG-22 (bottom). The left
                        column contains the dust-free CMDs; the middle panel contains
                        the CMDs computed using the extinction prescription
                        \ref{eqn:dust_extinct}; and the right column contains the CMDs
                        computed for three times the nominal amount of extinction. As
                        explained in Sect. \ref{ssec:isochrones}, sharp features are
                        visible in the CMDs due to each star particle constituting a
                        single stellar population, but this does not affect the analysis.}
        \label{fig:dustyCMDs}
\end{figure*}
 
We explored the effects of dust extinction on the resulting
  CMD and hence on the SFHs derived from it. Since
  dust is not explicitly included in the simulations, we incorporated
  dust extinction in post-processing following the procedure in
  \cite{Hopkins2005}, which relates dust extinction in the B-band to
  metallicity and the H{\sc I} column density:
\begin{equation} 
\frac{A_{B}}{N_{\rm H}} =
\frac{Z}{0.02}\ \Bigg(\frac{A_{B}}{N_{\rm H}}\Bigg)_{\rm MW}\label{eqn:dust_extinct}
, \end{equation} 
where $A_{B}$ is the B-band extinction, $N_{\rm H}$ is the hydrogen
column density, $Z$ is the gas metallicity, and $(A_{B}/N_{H})_{\rm
  MW}\ =\ \mathrm{8.47 \times 10^{-22}\ cm^{2}}$. We calculated the
hydrogen column density at the position of each of the star particles,
and, using the above equation, we get the extinction $A_{B}$ in the
B-band. Following \cite{Pei1992}, the extinction in the V- and I-bands
can be written in terms of $A_{B}$ as
\begin{align}
  A_{V} &= 0.78\ A_{B}, \\
A_I &= 0.44\ A_{B}.
\end{align}
To study the effects of dust extinction on the resultant SFH, the
extinction values thus calculated were included in the mock CMDs. The
result can be seen in Fig. \ref{fig:dustyCMDs}, where we show the
effect of the adopted dust prescription on the CMDs of DG-5 and DG-22.

\begin{figure}[h]
        \centering
        \includegraphics[width=\hsize]{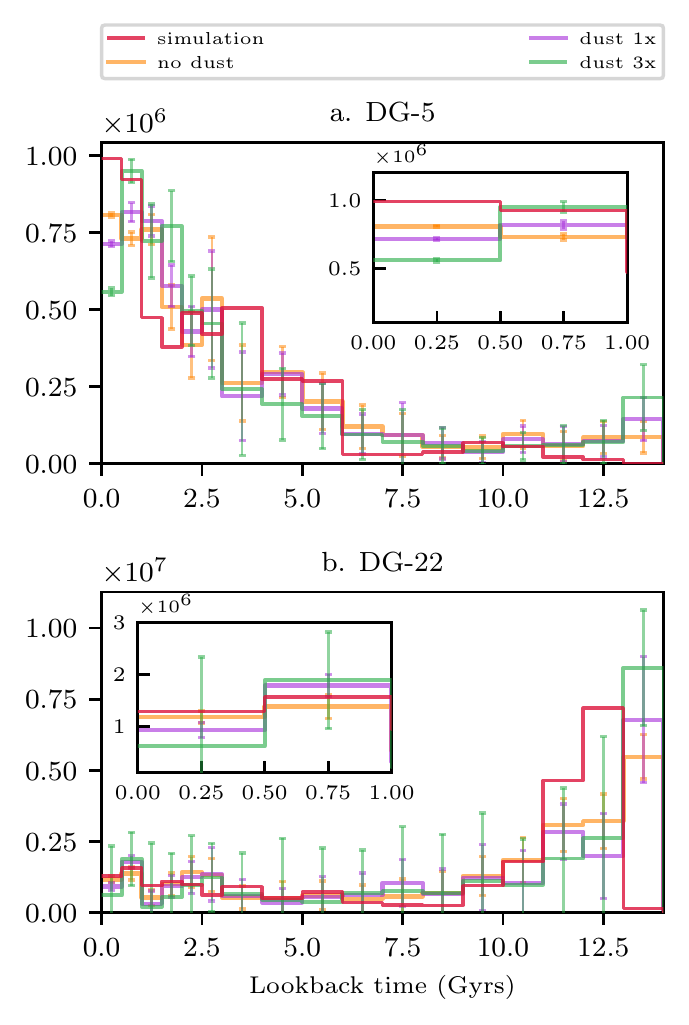}
        \caption{ Effect of dust absorption on the retrieved
                        SFR of DG-5 (top) and DG-22 (bottom). In both panels,
                the orange line represents SFR without taking dust
                extinction into account; the purple and green lines
                represent cases with varying amounts of dust extinction as
                indicated in the legend; and the red line represents the
                true SFR from the simulation snapshot data. The smaller panes show
        a zoomed-in view of the SFR in the most recent age bins, where the effects of dust extinction are most striking.}
        \label{fig:dust_obsc}
\end{figure}

The SFHs obtained from such dust-affected CMDs of DG-5 and
        DG-22 are shown in Fig. \ref{fig:dust_obsc}. These SFHs are
        compared to the case without dust extinction and the true SFH from
        the simulation star-particle data. The most striking effect of dust
        on the retrieved SFH is the significant underestimation of the SFR
        within the last 0.5~Gyr and the related overestimation of the SFR at
        larger lookback times, up to 1~Gyr. This can be explained as
        follows. Since young stellar particles preferentially reside in
        environments with high gas densities and metallicities, they will be
        the most strongly affected by dust extinction. This, to some extent,       depopulates the blue side of the main sequence while pushing stars
        toward fainter magnitudes and redder colors, leading to an
        underestimation of the most recent star formation and an
        overestimation of past star formation in the retrieved SFH. In DG-5,
        with a maximum extinction of 0.1~mag in the I-band, the SFR in the youngest age
        bin is underestimated by~15~\%. Likewise, in DG-22, with a maximum extinction of 0.8 mags in the I-band, the SFR in the youngest age bin is underestimated by ~25\% compared with the case without any dust extinction. 

Given the finite resolution inherent to the SPH formalism,
        very high-density and strongly obscured regions are absent from
        simulations such as these, and we may be underestimating the
        ``true'' extinction. To assess how strongly the
        presence of the amount of dust affects the retrieved SFH, we
        scaled up the nominal extinction by a factor of three for both
        galaxies. These CMDs are presented in the rightmost columns of
        Fig. \ref{fig:dustyCMDs}. The retrieved SFHs are shown in
        Fig. \ref{fig:dust_obsc}. The trend suggested by the fiducial dust
        extinction experiment is confirmed here:~The SFR in the most
        recent age bin is even more strongly underestimated (by $\sim
        40$~\% in the case of DG-5) while the SFR in older age bins, up to
        1~Gyr ago, is overestimated by $\sim 20$~\%. At even larger
        lookback times, the shuffling of stars leads to random variations
        of the retrieved SFR that stay well within the error bars.

In another, more extreme, test designed to mimic the very
        strong extinction of stars in high-density gas clouds, we simply
        removed all stellar particles from the CMD in regions where the
        gas density exceeds 1~amu~cm$^{-3}$. The SFH solved from such an
        extremely dust-affected CMD predictably shows a significantly
        lower recent SFR compared with the dust-free case
        (it is down by $\sim 30$~\%). Such lowered recent SFR is of course
        expected, as the new-born stars are obscured by the dense gas
        clouds in which they are formed. Since in this experiment we are
        simply removing stars and not reddening and dimming them, there is
        no accompanying overestimation of past star formation. In fact,
        the SFH is also lowered up to a few billion years of lookback
        time. This is most likely due to the ``accidental'' obscuration of
        older star particles that happen to reside inside a high-density
        gas cloud. This is not wholly unexpected since the star-formation
        regions are embedded in a background population of older stars.

\subsection{I$-$H versus I CMDs}

\begin{figure*}
        \centering
        \begin{subfigure}[t]{0.5\textwidth}
                \centering
                \caption*{(i) SFH of DG-5}
                \includegraphics{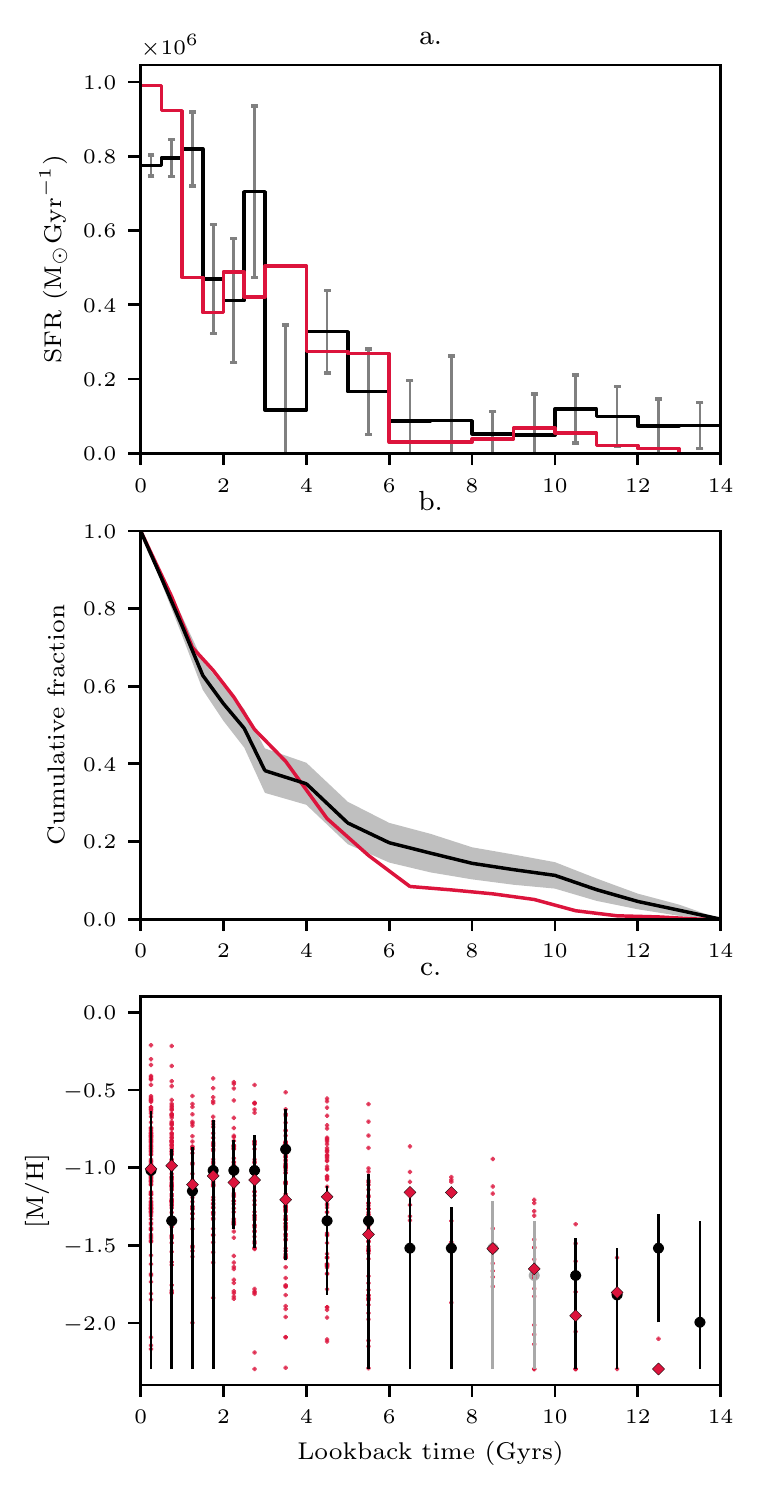}
        \end{subfigure}%
        ~
        \begin{subfigure}[t]{0.5\textwidth}
                \centering
                \caption*{(ii) SFH of DG-22}
                \includegraphics{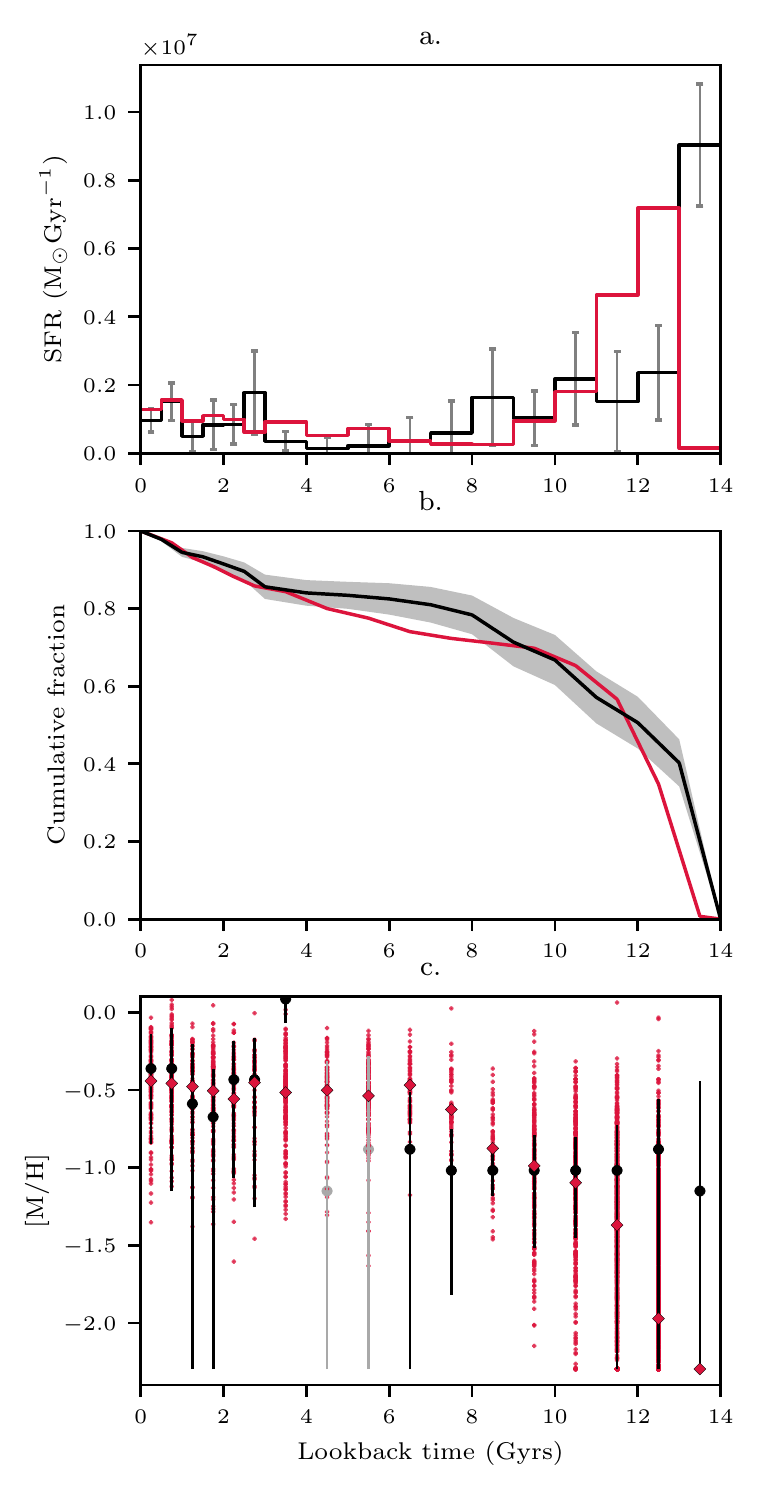}
        \end{subfigure}
        \caption{Results from the I, H CMD analysis of DG-5
            (left) and DG-22 (right): The panels and symbols have the
          same description as in Fig. \ref{fig:sfh_DG5_DG22_VI}.}
        \label{fig:sfh_DG5_DG22_IH}
\end{figure*}

With the focus of future instruments shifting more toward infrared
wavelengths, such as E-ELT/MICADO \citep{MICADO} and JWST/NIRCam
\citep{NIRCAM}, 
we also investigated the infrared (I, H) CMDs of
the simulated galaxies. Here, we show results from the analysis of
infrared CMDs of simulated dwarfs with three representative
SFH scenarios: (i) DG-5, with a recent
star formation; (ii) DG-18 (\ref{fig:IH1}), with an early
star-formation episode; and (iii) DG-20 (\ref{fig:IH1}), with a nearly
constant SFR. Results for DG-5 are shown in Fig.
\ref{fig:sfh_DG5_DG22_IH}, and those for DG-18 and DG-20 are shown in
Fig. \ref{fig:IH1}. We see that the I, H CMDs give results very
similar to those from the optical (V, I) CMDs and that the recovered
SFRs and AMRs adhere closely to those derived from the star-particle
data. Hence, SFH derived from resolved stellar
populations studies with the upcoming infrared instruments are also
directly comparable to those derived from galaxy simulations.

Due to the unavailability of crowding tables in the infrared band, the
observational errors in the I, H CMDs were simulated in a different
manner, as described below. The observational errors reported in
\cite{lcid3} (for the I-band) and in \cite{Dong2017} (for the H-band)
were approximated with polynomial functions of absolute
magnitudes. Using these relations, the corresponding values of errors
were obtained depending on the photometric magnitude of a star. These
values were then used as the standard deviations for Gaussian
distributions with zero mean, and the resultant magnitudes of a star
were obtained by convolving its magnitudes with random values
sampled from these Gaussians.


\section{Summary and conclusions} \label{sec:conclusions}

In this paper, we analyzed a set of 24 MoRIA dwarf galaxy
simulations using an observational approach to investigate
any systematic differences in the comparison of simulations with
observations. To do so, we created realistic V$-$I versus I CMDs of
simulated dwarf galaxies from their simulation star-particle
data. From these observed or mock CMDs, we derived the SFHs of the
simulated dwarf galaxies using the synthetic CMD method. The recovered
or solved SFHs were then compared to their true values from the
simulation star-particle data, mainly in terms of
the SFR and the AMR.

We find that the recovered SFHs are in very good agreement with the
true SFHs (Fig. \ref{fig:sfh_DG5_DG22_VI} and figures in Appendix
\ref{sec:AppA}). There are no systematic differences between
        the SFHs retrieved from the data and the star-particle data of
        the simulations, and we, therefore, conclude that quantities like the SFR
        and the AMR derived from the photometric observations of galaxies are
        directly comparable to their simulated counterparts.

Our experiments with dusty CMDs show that extinction and
        reddening can lead to a significant underestimation of the SFR
        during the most recent 0.5~Gyr, with the strength and the duration
        of this effect dependent on the amount of dust (quantified here by
        the maximum inflicted extinction). In turn, at larger lookback
        times, the SFR is overestimated.

As the focus of the next-generation instruments is shifting toward
the infrared range of the spectrum, 
we also analyzed the CMDs with
infrared bands (I- and H-bands). The synthetic CMD analysis of the
infrared I$-$H versus H CMDs gives results quite similar to those from the
optical V$-$I versus I CMDs, and in good agreement with the ground truth
from the simulation star-particle data (Fig.
\ref{fig:sfh_DG5_DG22_IH}). This paves the way for resolved stellar
population studies with future infrared facilities, such as JWST and
E-ELT.

\begin{acknowledgements}
We thank the anonymous referee for their valuable comments which have
helped improve the content of the manuscript. We would also like to
thank S. L. Hidalgo and I. Meschin for kindly providing the crowding
tables for our analysis.  SR, SDR and MM acknowledge financial support
from the European Union’s Horizon 2020 research and innovation
programme under Marie Skłodowska-Curie grant agreement No 721463 to
the SUNDIAL ITN network. This work has made use of the IAC-STAR
Synthetic CMD computation code.  IAC-STAR is supported and maintained
by the IAC's IT Division.

\\\\

\textit{Software:} \texttt{numpy} \citep{numpy}, \texttt{scipy}
\citep{scipy}, \texttt{matplotlib} \citep{matplotlib},
\texttt{astropy} \citep{astropy} and \texttt{pynbody} \citep{pynbody}.

\end{acknowledgements}

\bibliography{preprint_shivangee.bib}

\begin{thebibliography}{72}
\expandafter\ifx\csname natexlab\endcsname\relax\def\natexlab#1{#1}\fi

\bibitem[{{Aparicio} \& {Gallart}(2004)}]{iac-star}
{Aparicio}, A. \& {Gallart}, C. 2004, \aj, 128, 1465

\bibitem[{{Aparicio} \& {Hidalgo}(2009)}]{iac-pop}
{Aparicio}, A. \& {Hidalgo}, S.~L. 2009, \aj, 138, 558

\bibitem[{{Aparicio} {et~al.}(2016){Aparicio}, {Hidalgo}, {Skillman},
  {Cassisi}, {Mayer}, {Navarro}, {Cole}, {Gallart}, {Monelli}, {Weisz},
  {Bernard}, {Dolphin}, \& {Stetson}}]{2016ApJ...823....9A}
{Aparicio}, A., {Hidalgo}, S.~L., {Skillman}, E., {et~al.} 2016, \apj, 823, 9

\bibitem[{Battaglia {et~al.}(2012)Battaglia, Irwin, Tolstoy, de~Boer, \&
  Mateo}]{Battaglia2012}
Battaglia, G., Irwin, M., Tolstoy, E., de~Boer, T., \& Mateo, M. 2012, The
  Astrophysical Journal, 761, L31

\bibitem[{{Beichman} {et~al.}(2012){Beichman}, {Rieke}, {Eisenstein}, {Greene},
  {Krist}, {McCarthy}, {Meyer}, \& {Stansberry}}]{NIRCAM}
{Beichman}, C.~A., {Rieke}, M., {Eisenstein}, D., {et~al.} 2012, Society of
  Photo-Optical Instrumentation Engineers (SPIE) Conference Series, Vol. 8442,
  {Science opportunities with the near-IR camera (NIRCam) on the James Webb
  Space Telescope (JWST)}, 84422N

\bibitem[{Bernard {et~al.}(2012)Bernard, Ferguson, Barker, Hidalgo, Ibata,
  Irwin, Lewis, McConnachie, Monelli, \& Chapman}]{Bernard2012}
Bernard, E.~J., Ferguson, A.~M., Barker, M.~K., {et~al.} 2012, Monthly Notices
  of the Royal Astronomical Society, 420, 2625

\bibitem[{Bernard {et~al.}(2015)Bernard, Ferguson, Richardson, Irwin, Barker,
  Hidalgo, Aparicio, Chapman, Ibata, Lewis, McConnachie, \&
  Tanvir}]{Bernard2015a}
Bernard, E.~J., Ferguson, A. M.~N., Richardson, J.~C., {et~al.} 2015, Monthly
  Notices of the Royal Astronomical Society, 446, 2789

\bibitem[{{Bernard} {et~al.}(2018){Bernard}, {Schultheis}, {Di Matteo}, {Hill},
  {Haywood}, \& {Calamida}}]{bernard2018}
{Bernard}, E.~J., {Schultheis}, M., {Di Matteo}, P., {et~al.} 2018, \mnras,
  477, 3507

\bibitem[{{Bernstein} {et~al.}(2014){Bernstein}, {McCarthy}, {Raybould},
  {Bigelow}, {Bouchez}, {Filgueira}, {Jacoby}, {Johns}, {Sawyer}, {Shectman},
  \& {Sheehan}}]{2014SPIE.9145E..1CB}
{Bernstein}, R.~A., {McCarthy}, P.~J., {Raybould}, K., {et~al.} 2014, in
  \procspie, Vol. 9145, Ground-based and Airborne Telescopes V, 91451C

\bibitem[{{Boylan-Kolchin} {et~al.}(2011){Boylan-Kolchin}, {Bullock}, \&
  {Kaplinghat}}]{2011MNRAS.415L..40B}
{Boylan-Kolchin}, M., {Bullock}, J.~S., \& {Kaplinghat}, M. 2011, \mnras, 415,
  L40

\bibitem[{{Bullock} \& {Boylan-Kolchin}(2017)}]{2017ARA&A..55..343B}
{Bullock}, J.~S. \& {Boylan-Kolchin}, M. 2017, \araa, 55, 343

\bibitem[{{Cassisi} {et~al.}(2000){Cassisi}, {Castellani}, {Ciarcelluti},
  {Piotto}, \& {Zoccali}}]{bastiVLM}
{Cassisi}, S., {Castellani}, V., {Ciarcelluti}, P., {Piotto}, G., \& {Zoccali},
  M. 2000, \mnras, 315, 679

\bibitem[{{Castelli} \& {Kurucz}(2003)}]{Castelli&Kurucz2003}
{Castelli}, F. \& {Kurucz}, R.~L. 2003, in IAU Symposium, Vol. 210, Modelling
  of Stellar Atmospheres, ed. N.~{Piskunov}, W.~W. {Weiss}, \& D.~F. {Gray},
  A20

\bibitem[{{Chabrier}(2003)}]{chabrier2003}
{Chabrier}, G. 2003, \pasp, 115, 763

\bibitem[{{Cloet-Osselaer} {et~al.}(2012){Cloet-Osselaer}, {De Rijcke},
  {Schroyen}, \& {Dury}}]{2012MNRAS.423..735C}
{Cloet-Osselaer}, A., {De Rijcke}, S., {Schroyen}, J., \& {Dury}, V. 2012,
  \mnras, 423, 735

\bibitem[{{Cloet-Osselaer} {et~al.}(2014){Cloet-Osselaer}, {De Rijcke},
  {Vandenbroucke}, {Schroyen}, {Koleva}, \& {Verbeke}}]{2014MNRAS.442.2909C}
{Cloet-Osselaer}, A., {De Rijcke}, S., {Vandenbroucke}, B., {et~al.} 2014,
  \mnras, 442, 2909

\bibitem[{Cloet-Osselaer {et~al.}(2014)Cloet-Osselaer, {De Rijcke},
  Vandenbroucke, Schroyen, Koleva, \& Verbeke}]{Cloet-Osselaer2014}
Cloet-Osselaer, A., {De Rijcke}, S., Vandenbroucke, B., {et~al.} 2014, Monthly
  Notices of the Royal Astronomical Society, 442, 2909

\bibitem[{{Da Silva} {et~al.}(2012){Da Silva}, Fumagalli, \&
  Krumholz}]{DaSilva2012}
{Da Silva}, R.~L., Fumagalli, M., \& Krumholz, M. 2012, Astrophysical Journal,
  745

\bibitem[{{Davies} {et~al.}(2016){Davies}, {Schubert}, {Hartl}, {Alves},
  {Cl{\'e}net}, {Lang-Bardl}, {Nicklas}, {Pott}, {Ragazzoni}, {Tolstoy},
  {Agocs}, {Anwand-Heerwart}, {Barboza}, {Baudoz}, {Bender}, {Bizenberger},
  {Boccaletti}, {Boland }, {Bonifacio}, {Briegel}, {Buey}, {Chapron}, {Cohen},
  {Czoske}, {Dreizler}, {Falomo}, {Feautrier}, {F{\"o}rster Schreiber},
  {Gendron}, {Genzel}, {Gl{\"u}ck}, {Gratadour}, {Greimel}, {Grupp},
  {H{\"a}user}, {Haug}, {Hennawi}, {Hess}, {H{\"o}rmann}, {Hofferbert}, {Hopp},
  {Hubert}, {Ives}, {Kausch}, {Kerber}, {Kravcar}, {Kuijken}, {Lang-Bardl},
  {Leitzinger}, {Leschinski}, {Massari}, {Mei}, {Merlin}, {Mohr}, {Monna},
  {M{\"u}ller}, {Navarro}, {Plattner}, {Przybilla}, {Ramlau}, {Ramsay},
  {Ratzka}, {Rhode}, {Richter}, {Rix}, {Rodeghiero}, {Rohloff}, {Rousset},
  {Ruddenklau}, {Schaffenroth}, {Schlichter}, {Sevin}, {Stuik}, {Sturm},
  {Thomas}, {Tromp}, {Turatto}, {Verdoes-Kleijn}, {Vidal}, {Wagner}, {Wegner},
  {Zeilinger}, {Ziegler}, \& {Zins}}]{MICADO}
{Davies}, R., {Schubert}, J., {Hartl}, M., {et~al.} 2016, Society of
  Photo-Optical Instrumentation Engineers (SPIE) Conference Series, Vol. 9908,
  {MICADO: first light imager for the E-ELT}, 99081Z

\bibitem[{{de Boer} {et~al.}(2012){de Boer}, {Tolstoy}, {Hill}, {Saha},
  {Olszewski}, {Mateo}, {Starkenburg}, {Battaglia}, \& {Walker}}]{deBoer2012}
{de Boer}, T.~J.~L., {Tolstoy}, E., {Hill}, V., {et~al.} 2012, \aap, 544, A73

\bibitem[{{De Rijcke} {et~al.}(2013){De Rijcke}, Schroyen, Vandenbroucke,
  Jachowicz, Decroos, Cloet-Osselaer, \& Koleva}]{DeRijcke2013}
{De Rijcke}, S., Schroyen, J., Vandenbroucke, B., {et~al.} 2013, Monthly
  Notices of the Royal Astronomical Society, 433, 3005

\bibitem[{{Dolphin}(1997)}]{Dolphin1997}
{Dolphin}, A. 1997, \na, 2, 397

\bibitem[{Dolphin(2002)}]{Dolphin2002}
Dolphin, A.~E. 2002, Monthly Notices of the Royal Astronomical Society, 332, 91

\bibitem[{{Dong} {et~al.}(2017){Dong}, {Sch{\"o}del}, {Williams},
  {Nogueras-Lara}, {Gallego-Cano}, {Gallego-Calvente}, {Wang}, {Morris}, {Do},
  \& {Ghez}}]{Dong2017}
{Dong}, H., {Sch{\"o}del}, R., {Williams}, B.~F., {et~al.} 2017, \mnras, 470,
  3427

\bibitem[{{Fattahi} {et~al.}(2018){Fattahi}, {Navarro}, {Frenk}, {Oman},
  {Sawala}, \& {Schaller}}]{2018MNRAS.476.3816F}
{Fattahi}, A., {Navarro}, J.~F., {Frenk}, C.~S., {et~al.} 2018, \mnras, 476,
  3816

\bibitem[{{Ferguson} \& {Binggeli}(1994)}]{1994A&ARv...6...67F}
{Ferguson}, H.~C. \& {Binggeli}, B. 1994, \aapr, 6, 67

\bibitem[{{Gallart} {et~al.}(1996){Gallart}, {Aparicio}, \&
  {Vilchez}}]{Gallart1996a}
{Gallart}, C., {Aparicio}, A., \& {Vilchez}, J.~M. 1996, \aj, 112, 1928

\bibitem[{{Gardner} {et~al.}(2006){Gardner}, {Mather}, {Clampin}, {Doyon},
  {Greenhouse}, {Hammel}, {Hutchings}, {Jakobsen}, {Lilly}, {Long}, {Lunine},
  {McCaughrean}, {Mountain}, {Nella}, {Rieke}, {Rieke}, {Rix}, {Smith},
  {Sonneborn}, {Stiavelli}, {Stockman}, {Windhorst}, \&
  {Wright}}]{2006SSRv..123..485G}
{Gardner}, J.~P., {Mather}, J.~C., {Clampin}, M., {et~al.} 2006, \ssr, 123, 485

\bibitem[{{Gilmozzi} \& {Spyromilio}(2007)}]{EELT}
{Gilmozzi}, R. \& {Spyromilio}, J. 2007, The Messenger, 127

\bibitem[{{Grevesse} \& {Noels}(1993)}]{Grevesse&Noels1993}
{Grevesse}, N. \& {Noels}, A. 1993, in Origin and Evolution of the Elements,
  ed. N.~{Prantzos}, E.~{Vangioni-Flam}, \& M.~{Casse}, 15--25

\bibitem[{Haas \& Anders(2010)}]{Haas2010}
Haas, M.~R. \& Anders, P. 2010, Astronomy and Astrophysics, 512, A79

\bibitem[{Hidalgo {et~al.}(2011)Hidalgo, Aparicio, Skillman, Monelli, Gallart,
  Cole, Dolphin, Weisz, Bernard, Cassisi, Mayer, Stetson, Tolstoy, \&
  Ferguson}]{Hidalgo2011}
Hidalgo, S.~L., Aparicio, A., Skillman, E., {et~al.} 2011, Astrophysical
  Journal, 730 [\eprint{1101.5762}]

\bibitem[{Hopkins {et~al.}(2005)Hopkins, Hernquist, Martini, Cox, Robertson,
  Matteo, \& Springel}]{Hopkins2005}
Hopkins, P.~F., Hernquist, L., Martini, P., {et~al.} 2005, The Astrophysical
  Journal, 2003

\bibitem[{Hunter(2007)}]{matplotlib}
Hunter, J.~D. 2007, Computing in Science \& Engineering, 9, 90

\bibitem[{{Kauffmann} {et~al.}(1993){Kauffmann}, {White}, \&
  {Guiderdoni}}]{1993MNRAS.264..201K}
{Kauffmann}, G., {White}, S.~D.~M., \& {Guiderdoni}, B. 1993, \mnras, 264, 201

\bibitem[{{Klypin} {et~al.}(1999){Klypin}, {Kravtsov}, {Valenzuela}, \&
  {Prada}}]{1999ApJ...522...82K}
{Klypin}, A., {Kravtsov}, A.~V., {Valenzuela}, O., \& {Prada}, F. 1999, \apj,
  522, 82

\bibitem[{{McQuinn} {et~al.}(2015){McQuinn}, {Skillman}, {Dolphin}, {Cannon},
  {Salzer}, {Rhode}, {Adams}, {Berg}, {Giovanelli}, {Girardi}, \&
  {Haynes}}]{2015ApJ...812..158M}
{McQuinn}, K.~B.~W., {Skillman}, E.~D., {Dolphin}, A., {et~al.} 2015, \apj,
  812, 158

\bibitem[{{Meschin} {et~al.}(2014){Meschin}, {Gallart}, {Aparicio}, {Hidalgo},
  {Monelli}, {Stetson}, \& {Carrera}}]{Meschin2014}
{Meschin}, I., {Gallart}, C., {Aparicio}, A., {et~al.} 2014, \mnras, 438, 1067

\bibitem[{{Monelli} {et~al.}(2010){Monelli}, {Hidalgo}, {Stetson}, {Aparicio},
  {Gallart}, {Dolphin}, {Cole}, {Weisz}, {Skillman}, {Bernard}, {Mayer},
  {Navarro}, {Cassisi}, {Drozdovsky}, \& {Tolstoy}}]{lcid3}
{Monelli}, M., {Hidalgo}, S.~L., {Stetson}, P.~B., {et~al.} 2010, \apj, 720,
  1225

\bibitem[{{Moore} {et~al.}(1999){Moore}, {Ghigna}, {Governato}, {Lake},
  {Quinn}, {Stadel}, \& {Tozzi}}]{1999ApJ...524L..19M}
{Moore}, B., {Ghigna}, S., {Governato}, F., {et~al.} 1999, \apjl, 524, L19

\bibitem[{Oliphant(2006)}]{numpy}
Oliphant, T.~E. 2006, A guide to NumPy, Vol.~1 (Trelgol Publishing USA)

\bibitem[{{Papastergis} \& {Shankar}(2016)}]{2016A&A...591A..58P}
{Papastergis}, E. \& {Shankar}, F. 2016, \aap, 591, A58

\bibitem[{Pei(1992)}]{Pei1992}
Pei, Y.~C. 1992, The Astrophysical Journal, 395, 130

\bibitem[{{Pietrinferni} {et~al.}(2004){Pietrinferni}, {Cassisi}, {Salaris}, \&
  {Castelli}}]{Pietrinferni2004}
{Pietrinferni}, A., {Cassisi}, S., {Salaris}, M., \& {Castelli}, F. 2004, \apj,
  612, 168

\bibitem[{{Pietrinferni} {et~al.}(2006){Pietrinferni}, {Cassisi}, {Salaris}, \&
  {Castelli}}]{basti2006}
{Pietrinferni}, A., {Cassisi}, S., {Salaris}, M., \& {Castelli}, F. 2006, \apj,
  642, 797

\bibitem[{{Pietrinferni} {et~al.}(2013){Pietrinferni}, {Cassisi}, {Salaris}, \&
  {Hidalgo}}]{basti2013}
{Pietrinferni}, A., {Cassisi}, S., {Salaris}, M., \& {Hidalgo}, S. 2013, \aap,
  558, A46

\bibitem[{{Pineda} {et~al.}(2017){Pineda}, {Hayward}, {Springel}, \& {Mendes de
  Oliveira}}]{2017MNRAS.466...63P}
{Pineda}, J.~C.~B., {Hayward}, C.~C., {Springel}, V., \& {Mendes de Oliveira},
  C. 2017, \mnras, 466, 63

\bibitem[{{Planck Collaboration} {et~al.}(2016){Planck Collaboration}, {Ade},
  {Aghanim}, {Arnaud}, {Ashdown}, {Aumont}, {Baccigalupi}, {Banday},
  {Barreiro}, {Bartlett}, \& et~al.}]{2016A&A...594A..13P}
{Planck Collaboration}, {Ade}, P.~A.~R., {Aghanim}, N., {et~al.} 2016, \aap,
  594, A13

\bibitem[{{Pontzen} {et~al.}(2013){Pontzen}, {Ro{\v{s}}kar}, {Stinson}, \&
  {Woods}}]{pynbody}
{Pontzen}, A., {Ro{\v{s}}kar}, R., {Stinson}, G., \& {Woods}, R. 2013,
  {pynbody: N-Body/SPH analysis for python}

\bibitem[{{Price-Whelan} {et~al.}(2018){Price-Whelan}, {Sip{\H{o}}cz},
  {G{\"u}nther}, {Lim}, {Crawford}, {Conseil}, {Shupe}, {Craig}, {Dencheva},
  {Ginsburg}, {VanderPlas}, {Bradley}, {P{\'e}rez-Su{\'a}rez}, {de Val-Borro},
  {Paper Contributors}, {Aldcroft}, {Cruz}, {Robitaille}, {Tollerud},
  {Coordination Committee}, {Ardelean}, {Babej}, {Bach}, {Bachetti}, {Bakanov},
  {Bamford}, {Barentsen}, {Barmby}, {Baumbach}, {Berry}, {Biscani}, {Boquien},
  {Bostroem}, {Bouma}, {Brammer}, {Bray}, {Breytenbach}, {Buddelmeijer},
  {Burke}, {Calderone}, {Cano Rodr{\'\i}guez}, {Cara}, {Cardoso}, {Cheedella},
  {Copin}, {Corrales}, {Crichton}, {D{\textquoteright}Avella}, {Deil},
  {Depagne}, {Dietrich}, {Donath}, {Droettboom}, {Earl}, {Erben}, {Fabbro},
  {Ferreira}, {Finethy}, {Fox}, {Garrison}, {Gibbons}, {Goldstein}, {Gommers},
  {Greco}, {Greenfield}, {Groener}, {Grollier}, {Hagen}, {Hirst}, {Homeier},
  {Horton}, {Hosseinzadeh}, {Hu}, {Hunkeler}, {Ivezi{\'c}}, {Jain}, {Jenness},
  {Kanarek}, {Kendrew}, {Kern}, {Kerzendorf}, {Khvalko}, {King}, {Kirkby},
  {Kulkarni}, {Kumar}, {Lee}, {Lenz}, {Littlefair}, {Ma}, {Macleod},
  {Mastropietro}, {McCully}, {Montagnac}, {Morris}, {Mueller}, {Mumford},
  {Muna}, {Murphy}, {Nelson}, {Nguyen}, {Ninan}, {N{\"o}the}, {Ogaz}, {Oh},
  {Parejko}, {Parley}, {Pascual}, {Patil}, {Patil}, {Plunkett}, {Prochaska},
  {Rastogi}, {Reddy Janga}, {Sabater}, {Sakurikar}, {Seifert}, {Sherbert},
  {Sherwood-Taylor}, {Shih}, {Sick}, {Silbiger}, {Singanamalla}, {Singer},
  {Sladen}, {Sooley}, {Sornarajah}, {Streicher}, {Teuben}, {Thomas},
  {Tremblay}, {Turner}, {Terr{\'o}n}, {van Kerkwijk}, {de la Vega}, {Watkins},
  {Weaver}, {Whitmore}, {Woillez}, {Zabalza}, \& {Contributors}}]{astropy}
{Price-Whelan}, A.~M., {Sip{\H{o}}cz}, B.~M., {G{\"u}nther}, H.~M., {et~al.}
  2018, \aj, 156, 123

\bibitem[{{Read} {et~al.}(2016){Read}, {Agertz}, \&
  {Collins}}]{2016MNRAS.459.2573R}
{Read}, J.~I., {Agertz}, O., \& {Collins}, M.~L.~M. 2016, \mnras, 459, 2573

\bibitem[{{Revaz} \& {Jablonka}(2012)}]{2012A&A...538A..82R}
{Revaz}, Y. \& {Jablonka}, P. 2012, \aap, 538, A82

\bibitem[{Rubele {et~al.}(2011)Rubele, Girardi, Kozhurina-Platais, Goudfrooij,
  \& Kerber}]{Rubele2011}
Rubele, S., Girardi, L., Kozhurina-Platais, V., Goudfrooij, P., \& Kerber, L.
  2011, Monthly Notices of the Royal Astronomical Society, 414, 2204

\bibitem[{Ruiz-Lara {et~al.}(2018)Ruiz-Lara, Gallart, Beasley, Monelli,
  Bernard, Battaglia, S{\'{a}}nchez-Bl{\'{a}}zquez, Florido, P{\'{e}}rez, \&
  Mart{\'{i}}n-Navarro}]{Ruiz-Lara2018}
Ruiz-Lara, T., Gallart, C., Beasley, M., {et~al.} 2018, Astronomy and
  Astrophysics, 617, 1

\bibitem[{{Sales} {et~al.}(2017){Sales}, {Navarro}, {Oman}, {Fattahi},
  {Ferrero}, {Abadi}, {Bower}, {Crain}, {Frenk}, {Sawala}, {Schaller},
  {Schaye}, {Theuns}, \& {White}}]{2017MNRAS.464.2419S}
{Sales}, L.~V., {Navarro}, J.~F., {Oman}, K., {et~al.} 2017, \mnras, 464, 2419

\bibitem[{{Sawala} {et~al.}(2016){Sawala}, {Frenk}, {Fattahi}, {Navarro},
  {Bower}, {Crain}, {Dalla Vecchia}, {Furlong}, {Helly}, {Jenkins}, {Oman},
  {Schaller}, {Schaye}, {Theuns}, {Trayford}, \& {White}}]{2016MNRAS.457.1931S}
{Sawala}, T., {Frenk}, C.~S., {Fattahi}, A., {et~al.} 2016, \mnras, 457, 1931

\bibitem[{{Schaller} {et~al.}(2015){Schaller}, {Frenk}, {Bower}, {Theuns},
  {Jenkins}, {Schaye}, {Crain}, {Furlong}, {Dalla Vecchia}, \&
  {McCarthy}}]{2015MNRAS.451.1247S}
{Schaller}, M., {Frenk}, C.~S., {Bower}, R.~G., {et~al.} 2015, \mnras, 451,
  1247

\bibitem[{{Schroyen} {et~al.}(2011){Schroyen}, {de Rijcke}, {Valcke},
  {Cloet-Osselaer}, \& {Dejonghe}}]{2011MNRAS.416..601S}
{Schroyen}, J., {de Rijcke}, S., {Valcke}, S., {Cloet-Osselaer}, A., \&
  {Dejonghe}, H. 2011, \mnras, 416, 601

\bibitem[{{Shen} {et~al.}(2014){Shen}, {Madau}, {Conroy}, {Governato}, \&
  {Mayer}}]{2014ApJ...792...99S}
{Shen}, S., {Madau}, P., {Conroy}, C., {Governato}, F., \& {Mayer}, L. 2014,
  \apj, 792, 99

\bibitem[{{Skidmore} {et~al.}(2015){Skidmore}, {TMT International Science
  Development Teams}, \& {Science Advisory Committee}}]{2015RAA....15.1945S}
{Skidmore}, W., {TMT International Science Development Teams}, \& {Science
  Advisory Committee}, T. 2015, Research in Astronomy and Astrophysics, 15,
  1945

\bibitem[{{Skillman} {et~al.}(2017){Skillman}, {Monelli}, {Weisz}, {Hidalgo},
  {Aparicio}, {Bernard}, {Boylan-Kolchin}, {Cassisi}, {Cole}, {Dolphin},
  {Ferguson}, {Gallart}, {Irwin}, {Martin}, {Mart{\'{\i}}nez-V{\'a}zquez},
  {Mayer}, {McConnachie}, {McQuinn}, {Navarro}, \&
  {Stetson}}]{2017ApJ...837..102S}
{Skillman}, E.~D., {Monelli}, M., {Weisz}, D.~R., {et~al.} 2017, \apj, 837, 102

\bibitem[{{Snyder} {et~al.}(2015){Snyder}, {Torrey}, {Lotz}, {Genel},
  {McBride}, {Vogelsberger}, {Pillepich}, {Nelson}, {Sales}, {Sijacki},
  {Hernquist}, \& {Springel}}]{2015MNRAS.454.1886S}
{Snyder}, G.~F., {Torrey}, P., {Lotz}, J.~M., {et~al.} 2015, \mnras, 454, 1886

\bibitem[{{Springel}(2005)}]{gadget2}
{Springel}, V. 2005, \mnras, 364, 1105

\bibitem[{{Tolstoy} \& {Saha}(1996)}]{Tolstoy&Saha1996}
{Tolstoy}, E. \& {Saha}, A. 1996, \apj, 462, 672

\bibitem[{{Tosi} {et~al.}(1991){Tosi}, {Greggio}, {Marconi}, \&
  {Focardi}}]{Tosi1991}
{Tosi}, M., {Greggio}, L., {Marconi}, G., \& {Focardi}, P. 1991, \aj, 102, 951

\bibitem[{{van den Bosch} \& {Swaters}(2001)}]{2001MNRAS.325.1017V}
{van den Bosch}, F.~C. \& {Swaters}, R.~A. 2001, \mnras, 325, 1017

\bibitem[{Vandenbroucke {et~al.}(2016)Vandenbroucke, Verbeke, \& {De
  Rijcke}}]{Vandenbroucke2016}
Vandenbroucke, B., Verbeke, R., \& {De Rijcke}, S. 2016, Monthly Notices of the
  Royal Astronomical Society, 458, 912

\bibitem[{{Verbeke} {et~al.}(2017){Verbeke}, {Papastergis}, {Ponomareva},
  {Rathi}, \& {De Rijcke}}]{V17}
{Verbeke}, R., {Papastergis}, E., {Ponomareva}, A.~A., {Rathi}, S., \& {De
  Rijcke}, S. 2017, \aap, 607, A13

\bibitem[{{Verbeke} {et~al.}(2015){Verbeke}, {Vandenbroucke}, \& {De
  Rijcke}}]{V15}
{Verbeke}, R., {Vandenbroucke}, B., \& {De Rijcke}, S. 2015, \apj, 815, 85

\bibitem[{{Virtanen} {et~al.}(2019){Virtanen}, {Gommers}, {Oliphant},
  {Haberland}, {Reddy}, {Cournapeau}, {Burovski}, {Peterson}, {Weckesser},
  {Bright}, {van der Walt}, {Brett}, {Wilson}, {Jarrod Millman}, {Mayorov},
  {Nelson}, {Jones}, {Kern}, {Larson}, {Carey}, {Polat}, {Feng}, {Moore}, {Vand
  erPlas}, {Laxalde}, {Perktold}, {Cimrman}, {Henriksen}, {Quintero}, {Harris},
  {Archibald}, {Ribeiro}, {Pedregosa}, {van Mulbregt}, \&
  {Contributors}}]{scipy}
{Virtanen}, P., {Gommers}, R., {Oliphant}, T.~E., {et~al.} 2019, arXiv
  e-prints, arXiv:1907.10121

\bibitem[{{Wang} {et~al.}(2015){Wang}, {Dutton}, {Stinson}, {Macci{\`o}},
  {Penzo}, {Kang}, {Keller}, \& {Wadsley}}]{2015MNRAS.454...83W}
{Wang}, L., {Dutton}, A.~A., {Stinson}, G.~S., {et~al.} 2015, \mnras, 454, 83

\bibitem[{{Weisz} {et~al.}(2014){Weisz}, {Dolphin}, {Skillman}, {Holtzman},
  {Gilbert}, {Dalcanton}, \& {Williams}}]{2014ApJ...789..148W}
{Weisz}, D.~R., {Dolphin}, A.~E., {Skillman}, E.~D., {et~al.} 2014, \apj, 789,
  148

\end{thebibliography}

\begin{appendix}


\section{Complete results from the V$-$I versus I CMDs} \label{sec:AppA}
\begin{figure*}[hbt]
  \resizebox{\hsize}{!}
            {\includegraphics{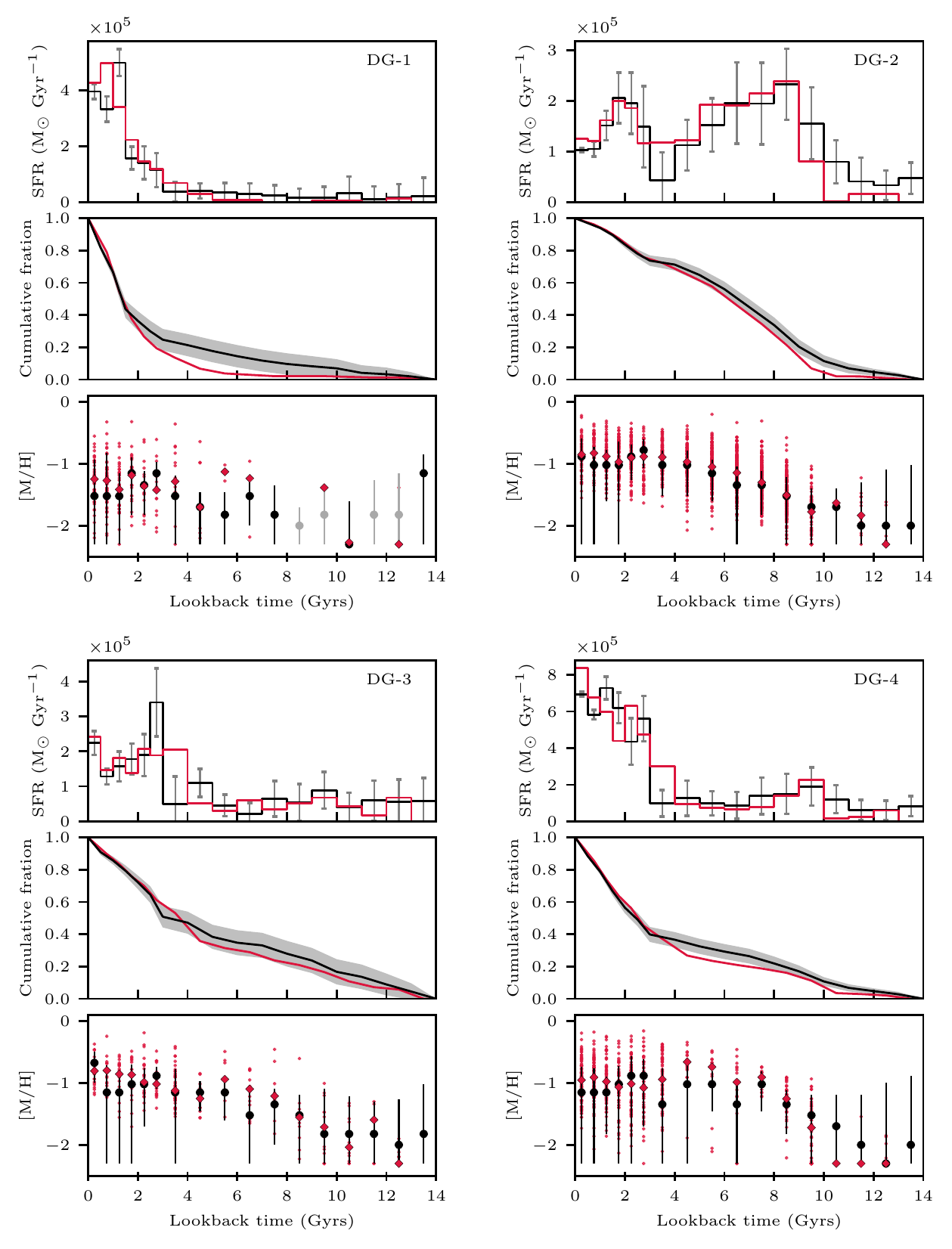}}
      \caption{Comparison of true SFH (in red) with that from the CMD
        fitting of mock CMDs (in black). The panels and symbols have
        the same description as in Fig. \ref{fig:sfh_DG5_DG22_VI}.}
        \label{fig:VI1}
\end{figure*}

\begin{figure*}
   \resizebox{\hsize}{!}
             {\includegraphics{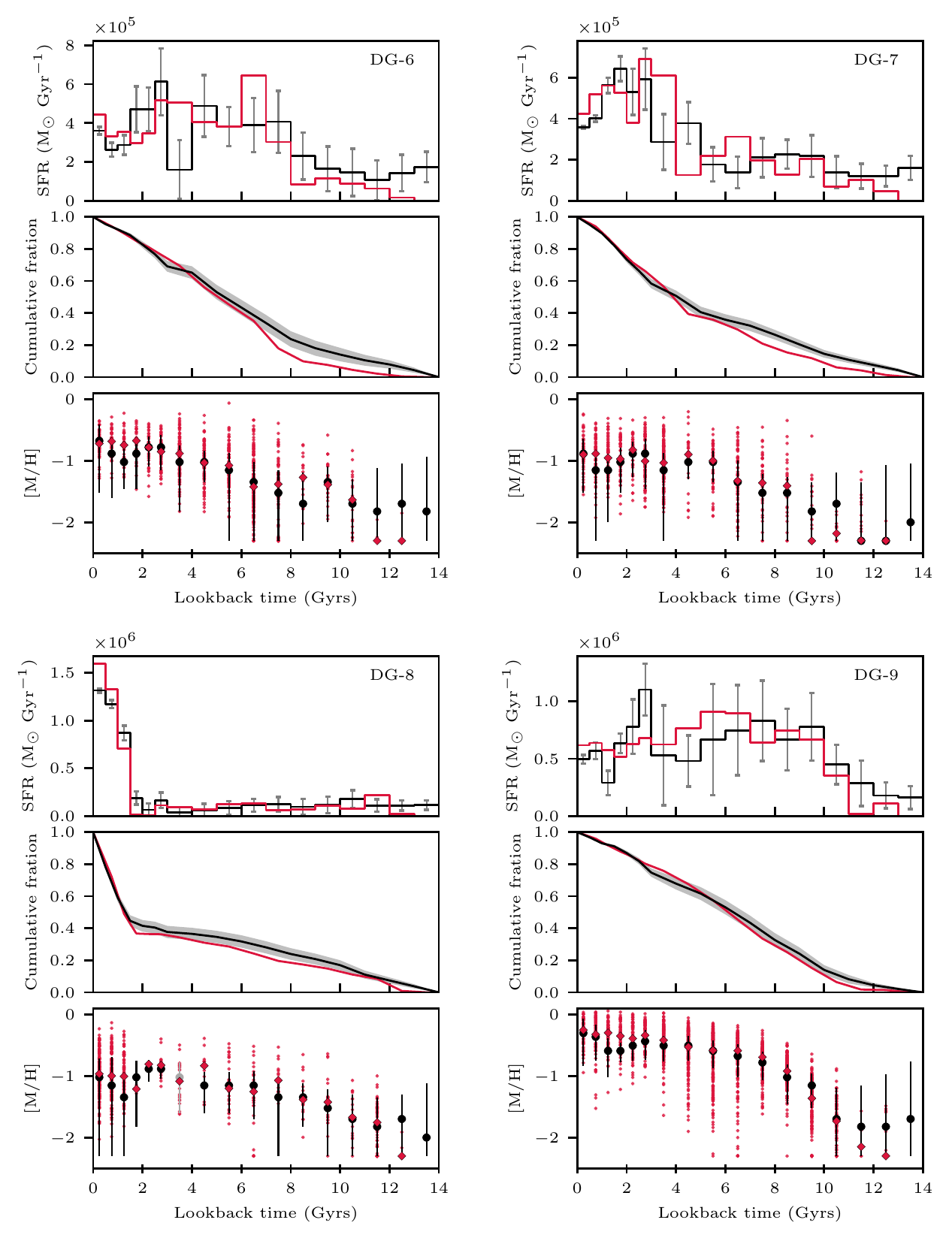}}
      \caption{Comparison of true SFH (in red) with that from the CMD
        fitting of mock CMDs (in black). The panels and symbols have
        the same description as in Fig. \ref{fig:sfh_DG5_DG22_VI}. }
        \label{fig:VI2}
\end{figure*}

\begin{figure*}
   \resizebox{\hsize}{!}
             {\includegraphics{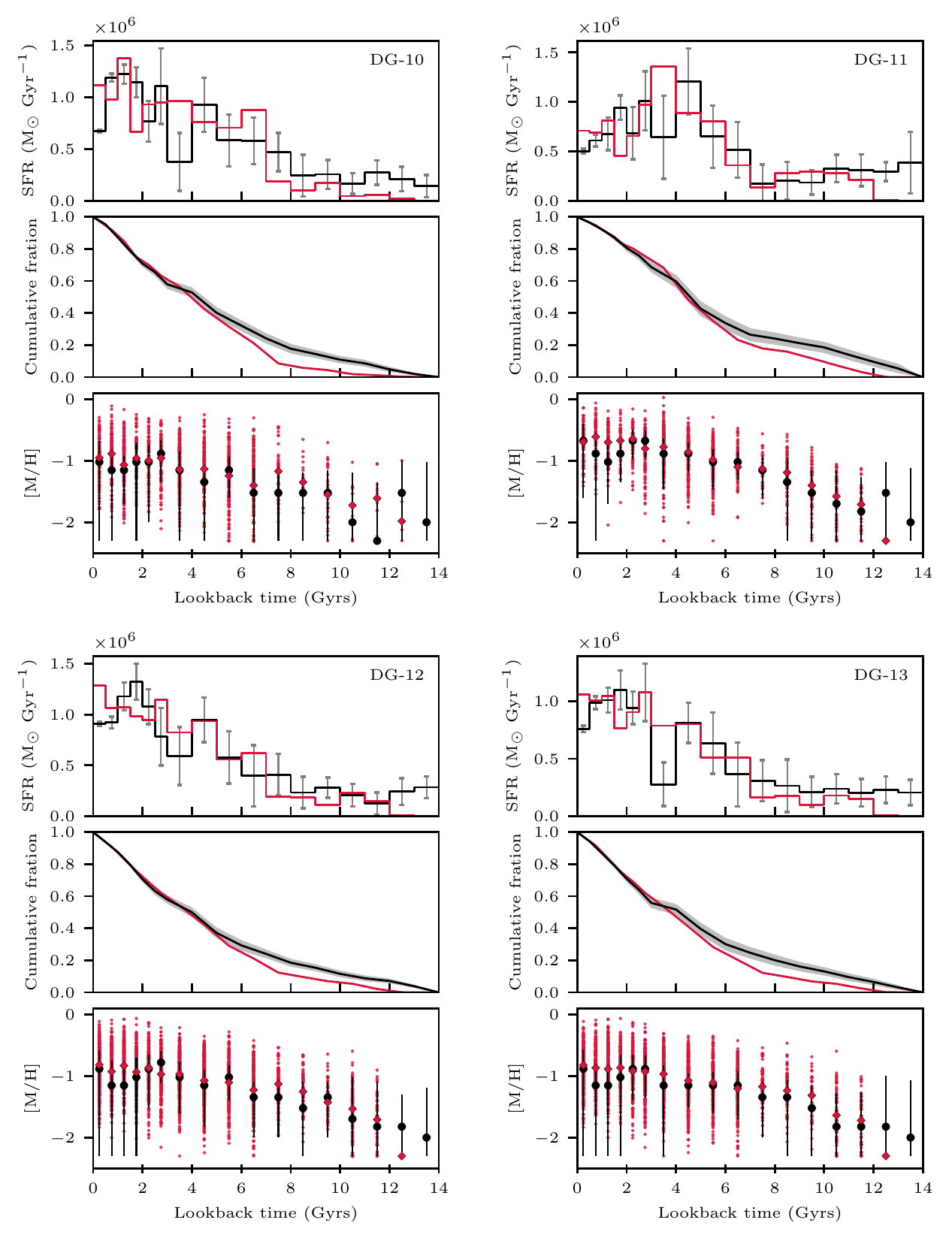}}
      \caption{Comparison of true SFH (in red) with that from the CMD
        fitting of mock CMDs (in black). The panels and symbols have
        the same description as in Fig. \ref{fig:sfh_DG5_DG22_VI}.}
        \label{fig:VI3}
\end{figure*}

\begin{figure*}
   \resizebox{\hsize}{!}
             {\includegraphics{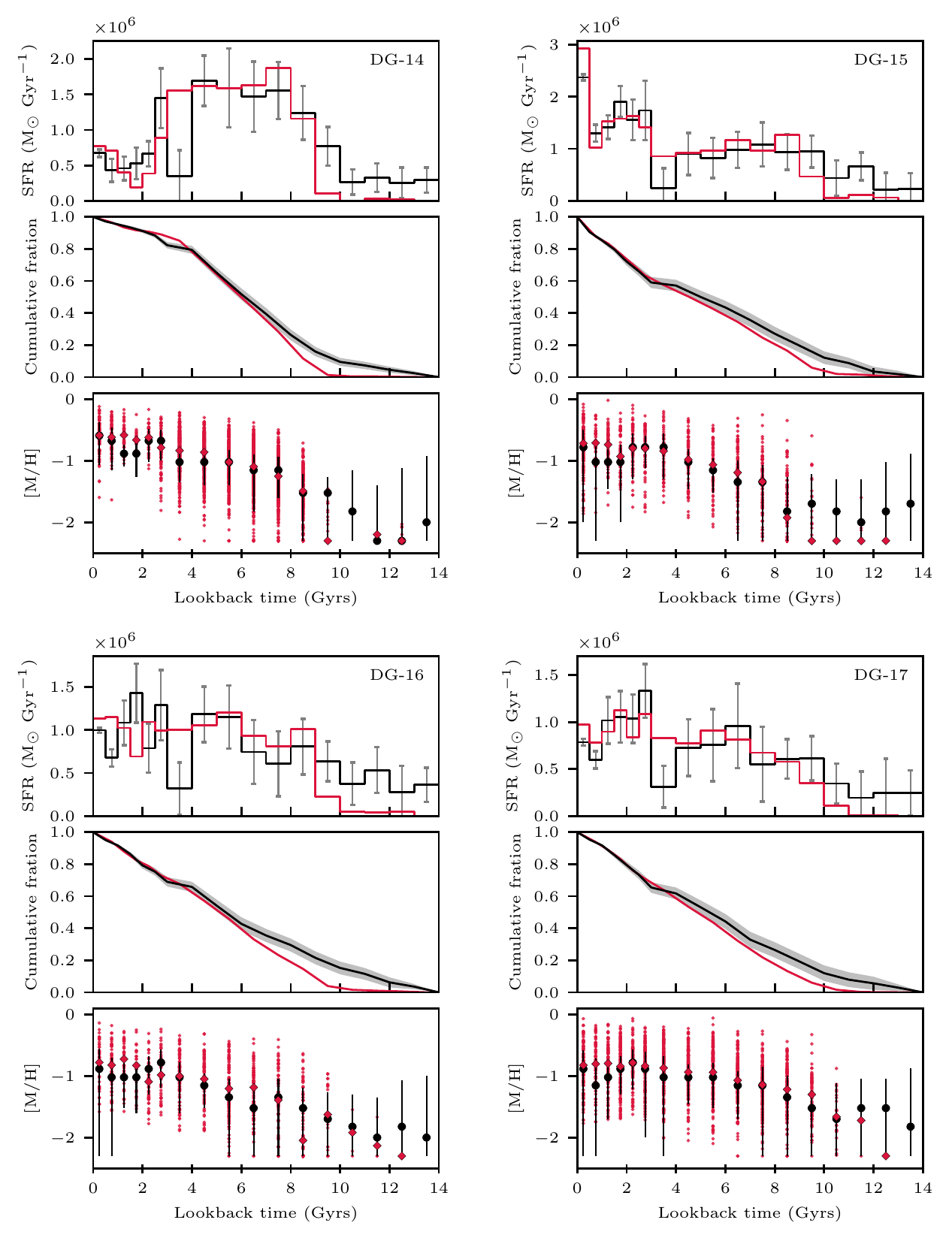}}
      \caption{Comparison of true SFH (in red) with that from the CMD
        fitting of mock CMDs (in black). The panels and symbols have
        the same description as in Fig. \ref{fig:sfh_DG5_DG22_VI}.}
        \label{fig:VI4}
\end{figure*}

\begin{figure*}
   \resizebox{\hsize}{!}
             {\includegraphics{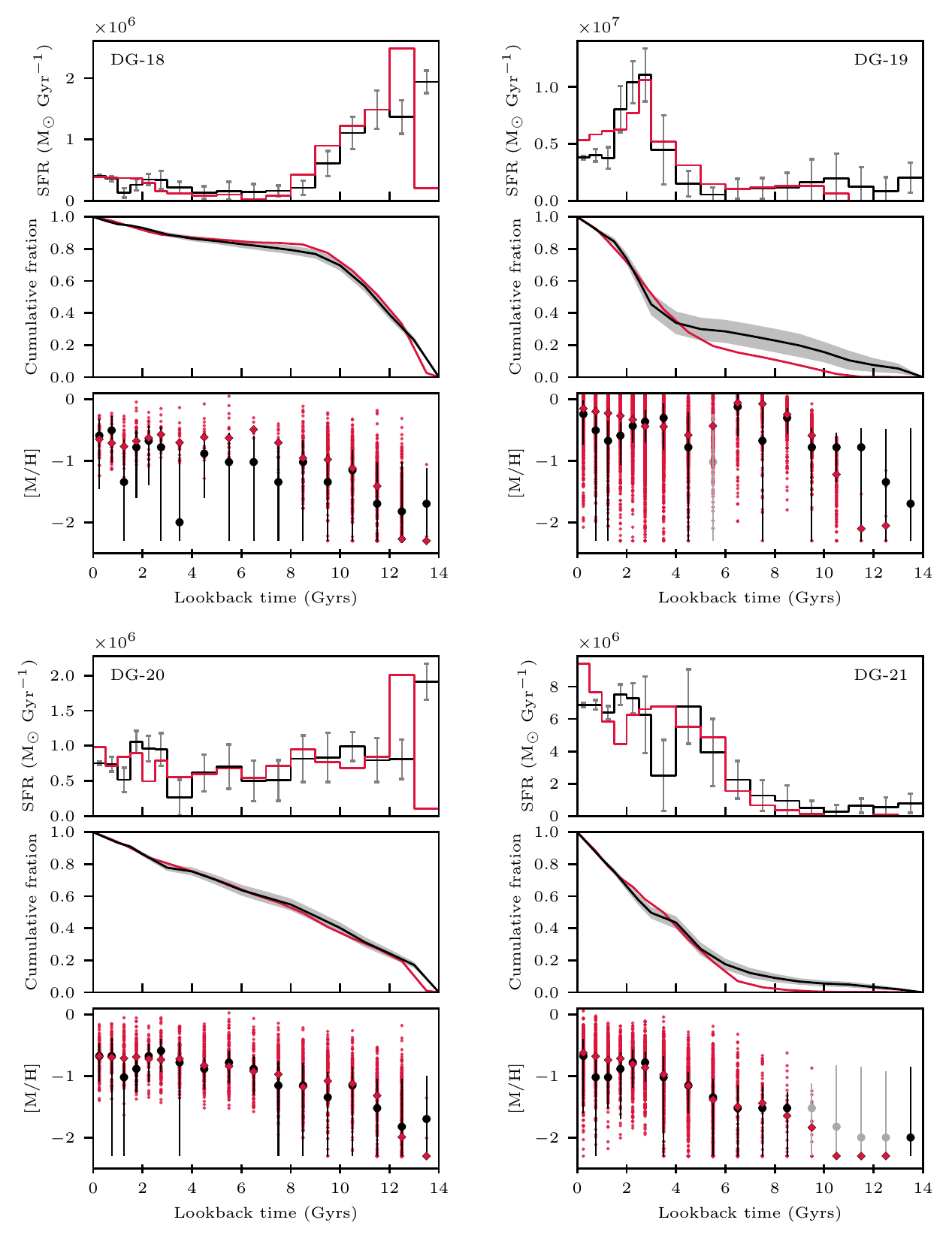}}
      \caption{Comparison of true SFH (in red) with that from the CMD
        fitting of mock CMDs (in black).The panels and symbols have
        the same description as in Fig. \ref{fig:sfh_DG5_DG22_VI}.}
        \label{fig:VI5}
\end{figure*}

\begin{figure*}
   \resizebox{\hsize}{!}
             {\includegraphics{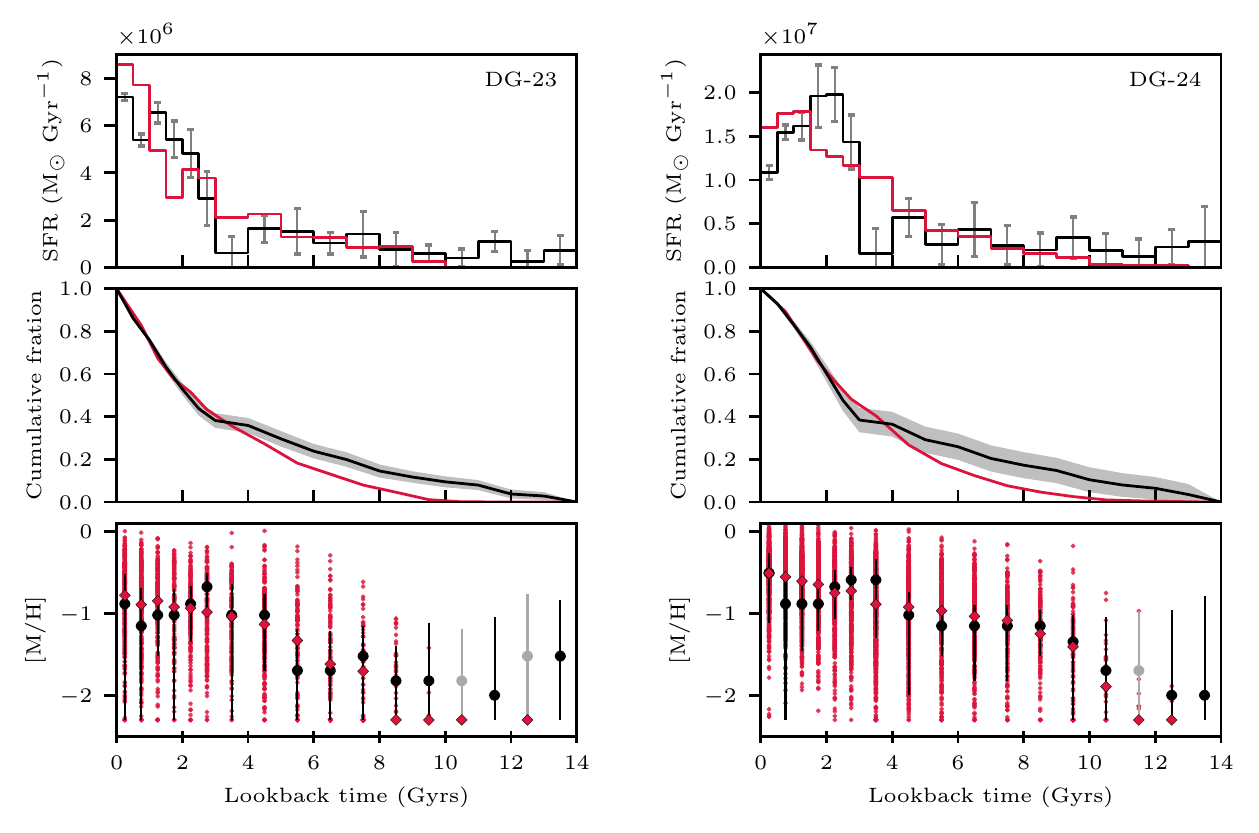}}
      \caption{Comparison of true SFH (in red) with that from the CMD
        fitting of mock CMDs (in black). The panels and symbols have
        the same description as in Fig. \ref{fig:sfh_DG5_DG22_VI}.}
        \label{fig:VI6}
\end{figure*}

Figures \ref{fig:VI1} to \ref{fig:VI6} show the comparison of the
solved SFHs with their true values for the complete set of simulations
studied in this work. They are arranged in increasing order of their
 total stellar mass.

\section{Results from the I, H CMDs (DG-18 and DG-20) } \label{sec:AppB}
\begin{figure*}
  \resizebox{\hsize}{!}  {\includegraphics{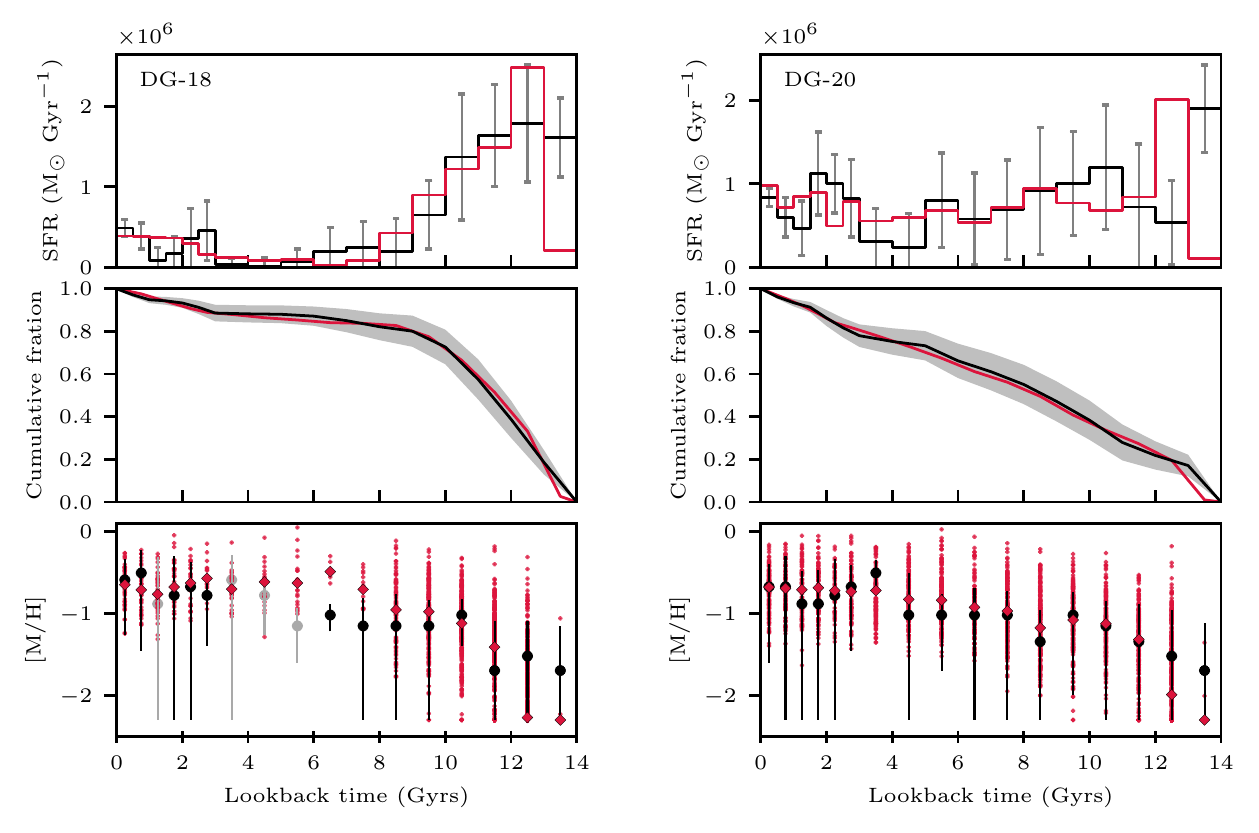}}
      \caption{Results from I, H CMD analysis of DG-18 and
          DG-20: comparison of true SFH (in red) with that from the
        CMD fitting of mock CMDs (in black). Panels and symbols have
        the same description as in Fig. \ref{fig:sfh_DG5_DG22_VI}. }
        \label{fig:IH1}
\end{figure*}

In view of the resolved stellar population studies with the next-generation infrared instruments on JWST \citep{NIRCAM} and E-ELT, we
performed the synthetic CMD analysis with the infrared CMDs, in
particular using the I- and H-bands. These bands are quite similar to
the proposed F090W- and F150W-bands for studying the resolved stellar
populations in some of the early release science of the JWST. Results
from the I, H CMD analysis of DG-18 (with an early star formation) and
DG-20 (with nearly constant star formation) are shown in Fig. \ref{fig:IH1}.

\end{appendix}

\end{document}